\documentclass[preprint,10pt]{elsarticle}

\usepackage[superscript,biblabel]{cite}
\usepackage{amsmath,amssymb,amsfonts}
\usepackage{algorithmic}
\usepackage{graphicx,xurl}
\usepackage{textcomp}
\usepackage{hyperref}
\usepackage[usenames]{xcolor}
\usepackage{comment}
\usepackage{caption}
\usepackage{soul}
\usepackage{subcaption}
\usepackage{paralist}
\usepackage{booktabs}
\usepackage{balance}
\usepackage{lastpage}
\usepackage{hanging}
\usepackage[multiple]{footmisc}

\makeatletter
\newcommand\footnoteref[1]{\protected@xdef\@thefnmark{\ref{#1}}\@footnotemark}
\makeatother

\newcommand*\ttvar[1]{\texttt{\expandafter\dottvar\detokenize{#1}\relax}}
\newcommand*\dottvar[1]{\ifx\relax#1\else
  \expandafter\ifx\string/#1\string/\allowbreak\else#1\fi
  \expandafter\dottvar\fi}



\usepackage{listings}
\definecolor{delim}{RGB}{20,105,176}
\definecolor{numb}{RGB}{106, 109, 32}
\definecolor{string}{rgb}{0.64,0.08,0.08}
\definecolor{lavender}{rgb}{0.45, 0.31, 0.59}

\lstdefinelanguage{json}{
    frame=tb,
    escapechar=\%,
    postbreak=\raisebox{0ex}[0ex][0ex]{\ensuremath{\color{gray}\hookrightarrow\space}},
    basicstyle=\ttfamily\footnotesize,
    upquote=true,
    morestring=[b]",
    stringstyle=\color{string},
    keywords={from,import},
    literate=
     *{0}{{{\color{numb}0}}}{1}
      {1}{{{\color{numb}1}}}{1}
      {2}{{{\color{numb}2}}}{1}
      {3}{{{\color{numb}3}}}{1}
      {4}{{{\color{numb}4}}}{1}
      {5}{{{\color{numb}5}}}{1}
      {6}{{{\color{numb}6}}}{1}
      {7}{{{\color{numb}7}}}{1}
      {8}{{{\color{numb}8}}}{1}
      {9}{{{\color{numb}9}}}{1}
      {\{}{{{\color{delim}{\{}}}}{1}
      {\}}{{{\color{delim}{\}}}}}{1}
      {[}{{{\color{delim}{[}}}}{1}
      {]}{{{\color{delim}{]}}}}{1},
}

\input{listings-python.prf}

\newif\iffinal
\finaltrue

\iffinal
  \newcommand\ian[1]{}
  \newcommand\ryan[1]{}
  \newcommand\ben[1]{}
  \newcommand\nicholas[1]{}
  \newcommand\kyle[1]{}
  \newcommand\mike[1]{}
  \newcommand\mikep[1]{}
  \newcommand\raf[1]{}
  \newcommand\nick[1]{}
  \newcommand\jim[1]{}
  \newcommand\zliu[1]{}
  \newcommand\alex[1]{}
  \newcommand\TODO[1]{}
\else
  \newcommand\ian[1]{\textcolor{red}{Ian: #1}} 
  \newcommand\ryan[1]{\textcolor{purple}{Ryan: #1}} 
  \newcommand\ben[1]{\textcolor{lavender}{Ben: #1}}
  \newcommand\nicholas[1]{\textcolor{orange}{Nicholas: #1}}
  \newcommand\kyle[1]{\textcolor{red}{Kyle: #1}}
  \newcommand\mike[1]{\textcolor{orange}{Mike: #1}}
  \newcommand\mikep[1]{\textcolor{orange}{MikeP: #1}}
  \newcommand\raf[1]{\textcolor{cyan}{Raf: #1}}
  \newcommand\nick[1]{\textcolor{green}{Nick: #1}}
  \newcommand\jim[1]{\textcolor{teal}{Jim: #1}}
  \newcommand\zliu[1]{\textcolor{green}{Zhengchun: #1}}
  \newcommand\alex[1]{\textcolor{teal}{Alex: #1}}
  \newcommand\TODO[1]{\textcolor{red}{TODO: #1}}
\fi

\usepackage{datenumber}
\newcounter{dateone}\newcounter{datetwo}%
\newcommand{\daydifftoday}[3]{%
  \setmydatenumber{dateone}{\the\year}{\the\month}{\the\day}%
  \setmydatenumber{datetwo}{#1}{#2}{#3}%
  \addtocounter{datetwo}{-\thedateone}%
  \thedatetwo
}

\begin{document}

\begin{frontmatter}

\title{Linking Scientific Instruments and Computation:\\Patterns, Technologies, Experiences}

\author[1]{Rafael Vescovi}
\author[1]{Ryan Chard}
\author[2]{Nickolaus D.\ Saint}
\author[1,2]{Ben Blaiszik}
\author[1,2]{Jim Pruyne} 
\author[1]{Tekin Bicer} 
\author[1]{Alex Lavens}
\author[1]{Zhengchun Liu}
\author[1,3]{Michael E.\ Papka}   
\author[1]{Suresh Narayanan}
\author[1]{Nicholas Schwarz}
\author[1,2]{Kyle Chard}
\author[1,2]{Ian T.\ Foster\corref{cor1}}

\cortext[cor1]{Corresponding author: foster@anl.gov, @ianfoster}

\affiliation[1]{organization={Argonne National Laboratory}, 
                addressline={9700 S Cass Ave},
                postcode={IL 60439}, city={Lemont}, country={USA}
                }

\affiliation[2]{organization={University of Chicago}, 
                addressline={5730 S Ellis Ave},
                postcode={IL 60615}, city={Chicago}, country={USA}
                }

\affiliation[3]{organization={University of Illinois Chicago}, 
                addressline={1200 W Harrison St},
                postcode={IL 60607}, city={Chicago}, country={USA}
                }

\end{frontmatter}

\section*{Summary} 
Powerful detectors at modern experimental facilities routinely collect data at multiple GB/s. Online analysis methods are needed to enable the collection of only interesting subsets of such massive data streams, such as by explicitly discarding some data elements or by directing instruments to relevant areas of experimental space. Thus, methods are required for configuring and running distributed computing pipelines---what we call flows---that link instruments, computers (e.g., for analysis, simulation, AI model training), edge computing (e.g., for analysis), data stores, metadata catalogs, and high-speed networks. We review common patterns associated with such flows and describe methods for instantiating these patterns. We present experiences with the application of these methods to the processing of data from five different scientific instruments, each of which engages powerful computers for data inversion, machine learning model training, or other purposes. We also discuss implications of such methods for operators and users of scientific facilities.

\section{Introduction}\label{sec:intro}

Humphry Davy observed that ``[n]othing tends so much to the advancement of knowledge as the application of a new instrument.''~\cite{davy}
Today, powerful new instruments such as upgraded synchrotron light sources~\cite{white2019new, apsu, daukantas2021synchrotron,chenevier2018esrf}, free-electron lasers~\cite{bostedt2016linac}, microscopes~\cite{eberle2018multi, bai2015cryo}, 
telescopes~\cite{andreoni2017deeper}, 
and robotic laboratories~\cite{flores2020materials, steiner2019organic, burger2020mobile} 
provide exciting new means to study phenomena in a broad range of scientific disciplines.

The power of these new instruments derives from their ability to probe reality rapidly and at fine spatial and temporal scales.
In so doing, they can generate data at rates (multi-GB/s) and volumes (100+ TB/day~\cite{wang2018synchrotron, rao2020synchrotrons}) that demand online computing, both to extract interesting information from data streams and to enable rapid configuration and steering of instruments to maximize information gained during scarce experimental time.
Tight coupling with powerful computing resources, such as data center clusters, high-performance computing (HPC), or AI accelerators, is often needed both to process this fire hose of data and to enable real-time feedback to experiments.

Such coupling requires flexible methods for coordinating actions and resources across diverse experimental and computing environments. 
We present common patterns for processing data from scientific instruments and describe tools that 
enable convenient specification of high-level \textit{flows} linking diverse actions and the flexible mapping of such flows onto diverse physical resources to meet reliability, scalability, timeliness, and security goals as an experiment runs.
(We use the term \textit{flow} rather than the over-used \textit{workflow} to emphasize our interest in capturing specialized data-processing patterns associated with scientific instrumentation.)
Specifically, we 
    (i) 
        identify common patterns encountered when scientists develop and run online data processing flows; 
    (ii) 
        show how Globus automation services~\cite{chard22automate, globusinstruments} can be used to capture such patterns;
    (iii)
        present experiences applying such methods in five different application scenarios; and
    (iv)
        examine the implications of such flows for both computing and experimental facilities.

\section{Patterns for Integration of Instruments and Computing}

Exponential growth in the rate at which instruments can perform measurements requires corresponding exponential improvements in the speed at which the resulting data are processed.
This means increasing use of automation and computation at every stage in the experimental process, including steps that were previously not rate-limiting and thus could be performed manually, such as recording and interpreting results and configuring the next experiment.  
New methods may be needed to capture data at high rates, extract interesting events in high rate streams, identify and filter out uninteresting phenomena, detect and/or correct errors, design further experiments, and perform simulations to choose between alternative experimental configurations---and to combine many such steps into automated experiment management \textit{flows}.


As in other areas of design, the identification of recurring patterns~\cite{Christopher77,gamma1995design} 
can contribute to cost reduction and performance improvement.
A design pattern captures a solution to a problem or class of problems in a re-usable form, via documentation of its purpose/intent, applicability, solution structure, and sample implementations.
In this section, we enumerate patterns we and others have observed when processing data from scientific instruments, and review the nature of the resources required to implement the patterns.

\subsection{What: Actions that are frequently included in flows} 


\textbf{Data collection}: 
  Capture data and associated metadata generated at high speeds, in unconventional formats, and on specialized devices, and make those data available to subsequent analyses. 

\textbf{Data reduction}:
  Reduce the volume of data to be processed and/or stored in other steps by applying either general-purpose compression~\cite{cappello2019use,vohl2017enabling} or  domain-specific feature detection (e.g., to find diffraction peaks in x-ray imaging~\cite{pokharel18hedm,liu22braggnn}). 

\begin{figure*}[ht]
     \centering
     
     \begin{subfigure}[b]{\textwidth}
         \centering
         \includegraphics[width=0.5\textwidth]{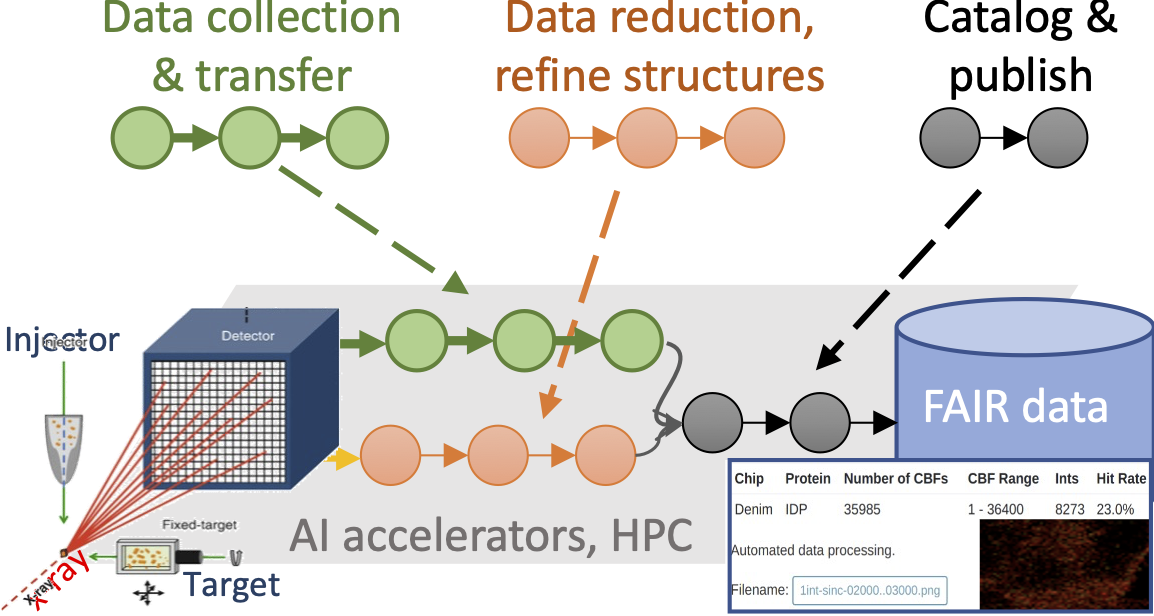}
         \caption{Serial synchrotron crystallography}
         \label{fig:example-ssx}
     \end{subfigure}
     
     \vspace{2ex}

     \begin{subfigure}[b]{\textwidth}
         \centering
         \includegraphics[width=0.5\textwidth]{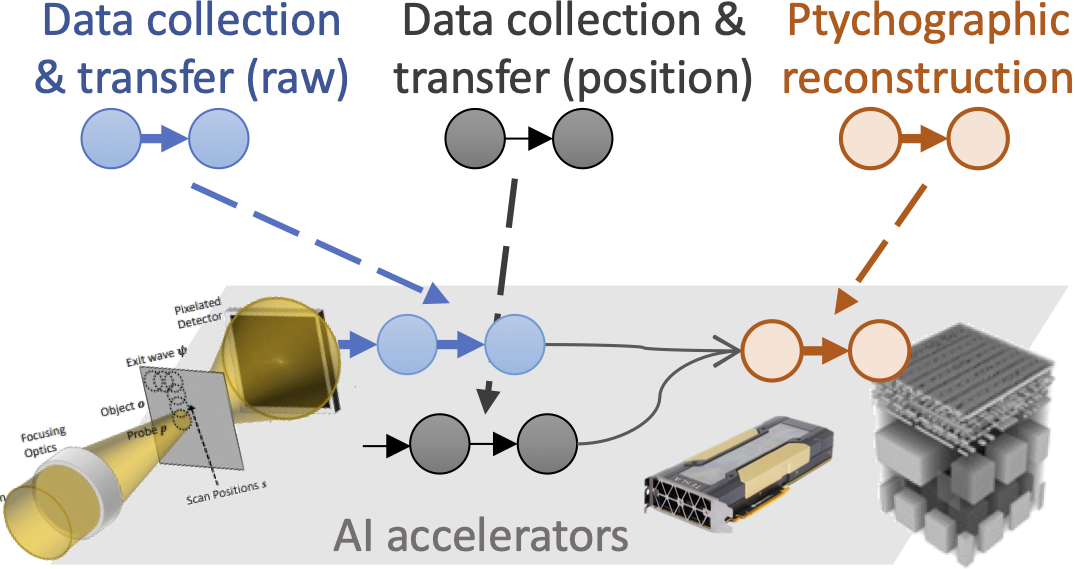}
         \caption{Ptychography}
         \label{fig:example-ptycho}
     \end{subfigure}
     
     \vspace{2ex}

     \begin{subfigure}[b]{\textwidth}
         \centering
         \includegraphics[width=0.5\textwidth]{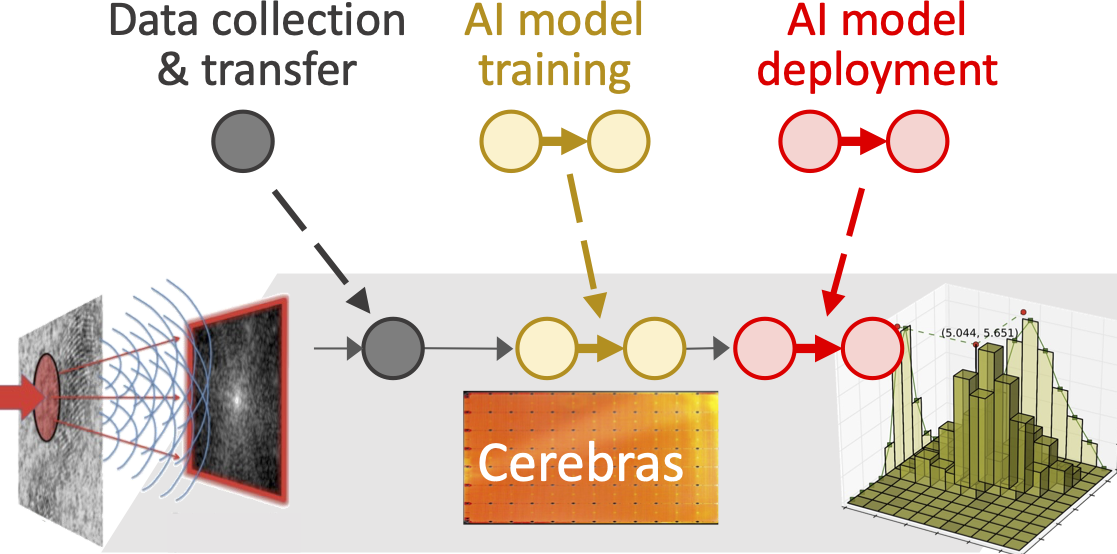}
         \caption{High energy diffraction microscopy}
         \label{fig:example-bragnn}
     \end{subfigure}
     
    \caption{The three subfigures show computing (e.g., data center clusters or AI accelerators, such as Cerebras) used to enable rapid collection and analysis of data from different classes of synchrotron light sources experiments. In each subfigure, we show, as directed acyclic graphs linking distinct actions, both the distinct flows used to automate different functions (above) and their deployment  deployed in the context of the applications (below).}
    \label{fig:example-flows}
\end{figure*}

\textbf{Data inversion}:
  Sophisticated computations are often required to convert sensor data into useful forms: for example, to generate a 3D or 4D representation from multiple 2D images~\cite{clackdoyle2010tomographic,nashed2014parallel}, or a 2D image from diffraction patterns. This step may be performed incrementally while data are collected or after all data are available.
  
\textbf{Machine learning (ML) model training}: 
  In this increasingly popular approach to data reduction, previously collected data (from current or prior experiments and/or simulations) are used to train ML models to recognize interesting phenomena for data reduction or rapid response~\cite{pelt2018improving,wasmer2018laser,liu2021bridging,li2021machine,konstantinova2022machine}.

\textbf{Experiment steering}: 
  Even better than discarding uninteresting data is to collect only interesting data in the first place.
  Scientists may use analyses of results from current or prior experiments to determine what experiment or measurement to perform next.
  Steering can range from fine-grained control of apparatus, such as taking (more) data from one part of a sample, to coarse-grained (e.g., choosing the next sample).
  Experiment steering can use design of experiment methods or more sophisticated
  active learning~\cite{kusne2020fly}, Gaussian processes~\cite{noack2021gaussian}, Bayesian optimization~\cite{zhang2022autonomous}, reinforcement learning~\cite{maffettone2021gaming}, or other methods.

\textbf{Coupled simulation}:
  Computational simulation can be used during experiment steering to eliminate (or prioritize) experimental configurations.

\textbf{Data storage and publication}:
  A flow may include steps to organize and store data and associated metadata (e.g., concerning experimental sample, configuration of apparatus, data processing steps) so as to make it findable, accessible, interoperable, and reusable (FAIR)~\cite{wilkinson2016}.

\subsection{Where: Alternative places to perform flow tasks} 

Analysis methods such as those just described can easily overwhelm instrument computers.
Indeed, some analyses can consume tens or even hundreds of thousands of cores~\cite{hidayetoglu2021memxct,mcclure2020toward}, albeit typically in a bursty manner. Similarly, experiments can generate petabytes.
The aggregate compute and storage demand across a research institution or multi-instrument research facility can be large, and shared (rather than per-instrument) computing facilities become attractive or even essential to exploit economies of scale in capital and operations costs.

Public cloud is a credible option for certain instrument workloads~\cite{chard18pct},
but data center systems can be more cost effective~\cite{cloudcosts}, especially when high-capacity, low-latency networks can support high data rate instruments and experiment steering.
Custom silicon may be required for certain data processing steps~\cite{bird2011computing,hammer2021strategies}.
Specialized accelerators may be used for tasks such as ML model training and inference~\cite{abeykoon2019scientific,chen2020survey,deiana2022applications}. 

When demand outstrips supply, adaptive methods may be used to direct compute and storage requests to different resources, prioritize certain tasks, and substitute alternative computational methods. 
In effect, computation may occur across a computing continuum~\cite{beckman2020harnessing,balouek2019towards,kumar2021coding} that extends from data acquisition computers co-located with experiments to powerful clusters in data centers. For a given flow, computation may occur at multiple points across this continuum. For example, rapid quality control may be executed near an instrument on a co-located device, machine learning training on specialized AI hardware, and large-scale reconstruction on a data center cluster. The ``best'' location for a computation can be hard to determine and may change over time according to data location, resource availability, cost, and performance.

\subsection{Example realizations of patterns}

The three flows in \autoref{fig:example-flows}, to be described in more detail in \autoref{sec:app-experiences}, illustrate some of the elements just described.
\textit{Serial synchrotron crystallography} (SSX) experiments collect diffraction data from target crystals. Several flows combine to process batches of acquired images, identify `hits,' refine crystal structures, and catalog results for later use. 
\textit{Ptychography} is a diffraction imaging technique that can produce images with high resolution. The flow shown here first transfers raw and position data to specialized compute resources before executing 2D reconstruction on GPUs.  
\textit{High energy diffraction microscopy} (HEDM) is used to characterize polycrystalline microstructures. This flow uses acquired data to train a neural network model for detecting peak positions in raw data. After training on a suitable AI accelerator, the flow transfers the trained model to the instrument for online use.

\begin{figure*}[h!]
\centering
\includegraphics[width=\textwidth]{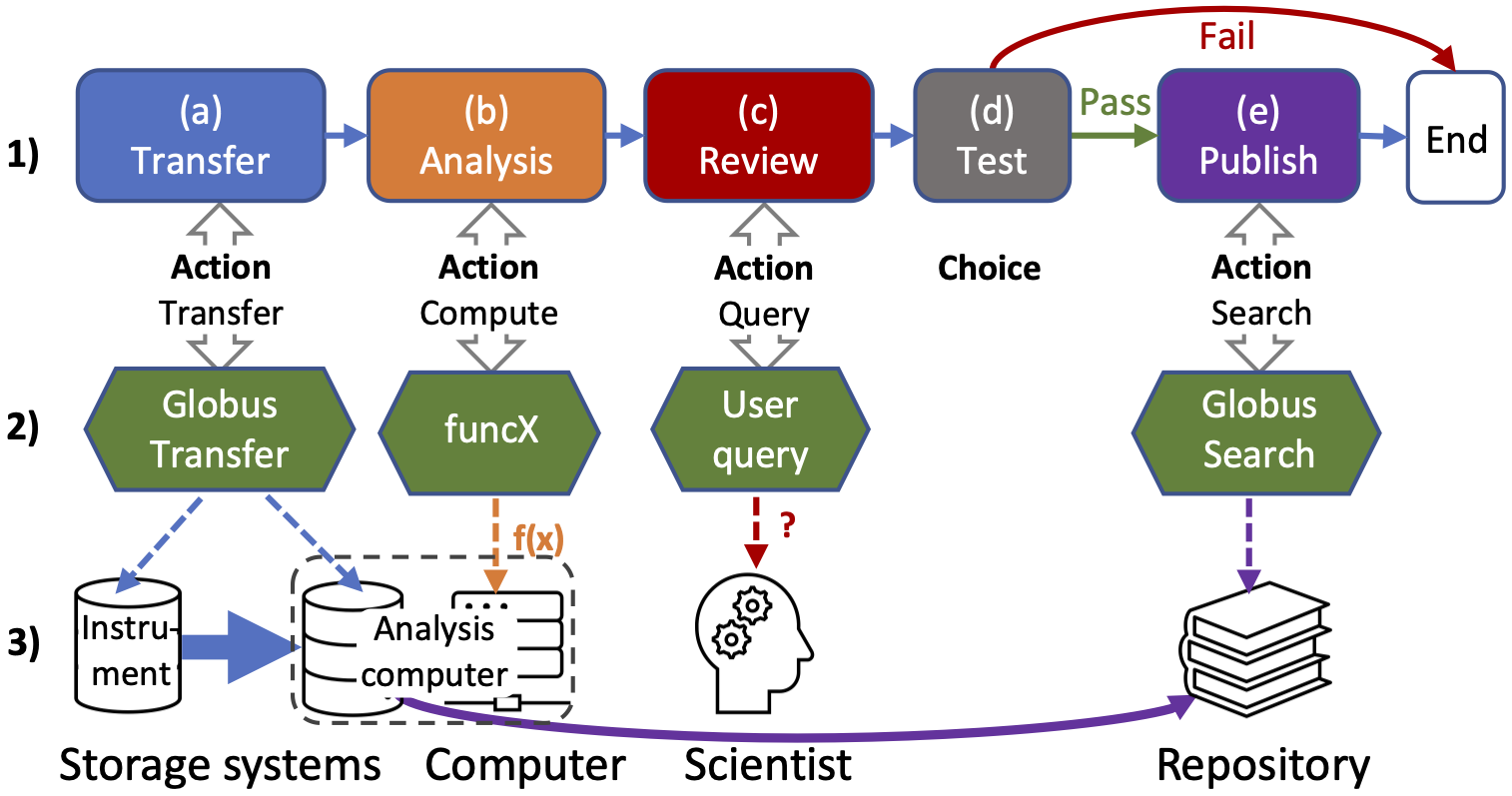}
\caption{From top to bottom: 
1) User perspective of a simple \textbf{flow} that, successively (shown left to right), 
	(a) transfers data from an instrument to an analysis computer;
	(b) runs an analysis; 
	(c) asks a user to review the analysis result; and, if
	(d) the user review is positive, 
	(e) publishes the data to a repository.
2) The Globus \textbf{platform services} engaged by the {\sc{Transfer}},  {\sc{Compute}}, {\sc{Query}}, and {\sc{Search}} action providers.
3) The \textbf{resources} interacted with by those platform services: instrument storage system; colocated analysis storage system and storage computer; scientist; and data repository.  
Not shown are the Globus Auth service that handles identities and access tokens, and the Globus Flows service that coordinates flow execution.
\label{fig:flow-flow}}
\end{figure*}

\section{Implementing Flows with the Globus Platform}

Having reviewed patterns for coupling experimental facilities with computation, we now examine how these patterns may be realized in practice, with the goal of providing actionable information that readers can apply to develop and execute their own flows. 

We believe strongly that the widespread integration of scientific instruments into computational flows requires \textbf{reusable flow specifications} that can be easily \textbf{adapted} to different applications, instruments, and computational environments.
Thus, our chosen approach to flow authoring and execution combines \textbf{automation services} for the specification and execution of flows with a \textbf{research automation fabric} to enable decoupling of abstract automation actions (e.g., move data, run program, publish records) from the specifics of individual data stores, computers, and catalogs---so that, for example, different compute and storage tasks can be directed to different resources (e.g., data center cluster, cloud, local accelerator), depending on needs and availability.
In the following, we describe these two sets of capabilities in turn.
For concreteness in presentation, we employ capabilities provided by the Globus platform~\cite{ananthakrishnan2015globus}.

\subsection{The Globus Research Automation Fabric}\label{sec:globus}

The Globus platform comprises a set of cloud-hosted services to which users can make various requests: for example, to transfer data from one storage system to another; run a computation on a computer; and load or search data in a catalog. In each case, the appropriate cloud service handles details such as authentication, authorization, monitoring of progress, and retries on failure that would otherwise hinder a scientist's work.
We leverage the following Globus services:
\begin{itemize}
    \item 
        \textbf{IAM services} (Auth, Groups) for single sign-on and management of identities and credentials, and delegation.
    \item 
        \textbf{Data services} (Transfer, HTTPS,
        Share) for access to, and managed movement of, files.
    \item
        \textbf{Metadata management} (Search, Identifiers) for indexing and generating persistent references to data.
    \item 
        \textbf{Compute services} (funcX, OAuthSSH) 
        for invocation and management of computational tasks.  
    \item 
        \textbf{Automation services} (Flows, Triggers, Queues) for execution of flows. 
\end{itemize}

The Globus Transfer~\cite{allen12software} and funcX~\cite{chard20funcx} services interact with local proxy agents deployed on storage systems and computers, respectively: Globus collections (implemented by Globus Connect software) for data actions, and funcX endpoints (implemented by funcX software) for compute actions.
These agents are deployed persistently at many experimental and computational facilities, and can also be deployed as needed by scientists.
The Globus cloud services plus the proxy agents implement a universal compute and data fabric that encompasses any and all resources on which agents are deployed---in aggregate, 10,000s of resources at 1000s of institutions worldwide, ranging from cloud providers to clusters, supercomputers, and AI accelerators.
Searchable registries support the discovery of agents that a user has permission to access.

All Globus platform services leverage the Globus Auth security fabric~\cite{tuecke16auth} for management of user identities and credentials, generation of OAuth~2 access tokens~\cite{oauth2} for programmatic invocation of services, and generation of delegation tokens that allow services to act on a user's behalf.
Crucially, data and computation remain at the edge: they never reach the cloud. Globus high assurance service levels allow for management of protected (e.g., HIPAA) data.

\subsection{Globus Automation Services}\label{sec:FA}

Globus automation services---Globus Flows, Triggers, and Queues~\cite{chard22automate}---build on the fabric provided by Globus platform services to allow scientists to specify and execute sequences of \textbf{actions} (or, sometimes, choices) called \textbf{flows}.  
A flow is specified as a JSON document---or, as described below, by using a Python toolkit, Gladier (for \textit{Globus Architecture for Data-Intensive Experimental Research}).
Flow execution may be invoked explicitly by the scientist, or triggered by an external event, such as generation of new data at an instrument. 
The Globus Flows service then manages flow execution.
A web interface allows users to monitor the progress of a flow's execution, and to detect and diagnose errors: see \S\ref{sec:flowsUI}.

\autoref{fig:flow-flow} shows an example flow and provides more details on how flows are implemented.
Each type of action that may be invoked in a flow is handled by a persistent \textbf{action provider} service. 
Action providers can run programs (funcX~\cite{chard20funcx}, OAuthSSH~\cite{alt20oauthssh}), transfer files (Globus Transfer~\cite{allen12software,liu2021design}), publish data to catalogs (Globus Search~\cite{ananthakrishnan18platform}), manage data permissions (Globus Share~\cite{chard2014efficient}), and generate persistent identifiers (Globus Identifiers~\cite{ananthakrishnan20identifiers}), among other tasks relevant to instrument data processing. 
In general, an action provider implements flow actions by requesting that the appropriate service (e.g., Globus Transfer, funcX) initiate the action, and then polling periodically to see whether the action has completed.
(As we discuss later, this polling can be a source of overhead.)
All action provider services implement a consistent, asynchronous REST API~\cite{fielding2000architectural}, facilitating the integration of new activities. 
Additional action providers may be deployed to support specific instruments, compute resources, or other customized needs by adhering to a well-defined interface~\cite{chard22automate}.


The implementation of Globus-operated action services, like those of other Globus platform services, leverages cloud services (e.g., Amazon Lambda, Step Functions, Simple Queue Service) for reliability and scalability.
Cloud-based hosting enables delivery of research process automation capabilities to a wide user base, without requiring users to download and install software. 
It also provides economies of scale, thereby reducing the costs associated with distributing software.


\subsection{The Gladier Toolkit}

We have developed a Python toolkit, Gladier~\cite{gladiersoftwarerepo}, to assist in the authoring and management of flows for instrument science.
This toolkit defines wrapper functions for registering funcX actions and flow definitions, 
invoking a new instance of a flow (a ``run'') with specified inputs,
and monitoring a specified directory for file events.
These functions allow for concise definitions of flows that integrate instrument and computation, as shown in \autoref{lst:simple_ssx}.

A Gladier user deploys client libraries on remote sources (e.g., on a computer co-located with an experiment) to detect events and invoke flows. 
A Gladier \textit{tool} definition, implemented as a Python object, provides the information needed to populate a flow action. The Gladier toolkit provides implementations of common tools (e.g., transfer) as well as examples for experiment-specific tools (e.g., Stills processing with the Diffraction Integration for Advanced Light Sources (DIALS) package~\cite{dials}); 
users may add other tools by implementing the Python class.
To deploy and run a flow, users simply provide a list of tools to be used along with specific flow input arguments. Gladier uses this specification to register the necessary funcX functions and create and then register the flow definition. 

We observe that flows for different experiments tend to follow similar patterns, independent of the experiment modality; the major area of customization concerns application-specific functions used to operate on data. Thus, we find that scientists often can employ an existing flow unchanged, simply specifying different compute and data endpoint identifiers and storage paths; different processing function(s); and a different Globus Search catalog for publication. 
In other cases, they can adapt an existing flow by adding and deleting tools from the description, and writing and deploying new funcX functions as required. Further, users can create, version, and share custom tools via GitHub, making them available for others to adopt within other flows.

The Gladier toolkit represents a relatively early attempt
to provide a Pythonic interface to Globus Flows. 
Experiences thus far have been positive.
Nevertheless, we imagine that future applications will motivate extensions---for example, to simplify specification of conditional execution and input schemas, both supported in Globus Flows but not handled well in the current toolkit.
We expect to develop other interfaces (e.g., web) to support other communities.

\begin{figure}[ht]
\begin{lstlisting}[language=json, label=lst:simple_ssx, 
caption={A simple SSX analysis flow, as defined with the Gladier toolkit.
The flow comprises two tasks, one for the Transfer from instrument to a compute resource, and one to run the DIALS stills processing function on the transferred data.
For brevity, we use \textbf{\emph{U1}}, \textbf{\emph{U2}}, and \textbf{\emph{U3}}, and
\textbf{\emph{P1}} and \textbf{\emph{P2}}, to represent UUIDs and paths, respectively.}]
from gladier import GladierBaseClient

@generate_flow_definition
class SSXFlow(GladierBaseClient):
  gladier_tools = [
    'gladier_tools.tools.Transfer',
    'gladier_ssx.tools.DialsStills'
  ]
  
flow_input = {
  'funcx_endpoint': %{\emph{\textbf{U1}}%,
  'transfer_source_endpoint_id': %{\emph{\textbf{U2}}%,
  'transfer_destination_endpoint_id': %{\emph{\textbf{U3}}%,
  'transfer_source_path': %{\emph{\textbf{P1}}%,
  'transfer_destination_path': %{\emph{\textbf{P2}}%,
}

ssx_flow_client = SSXFlow()
run_id = ssx_flow_client.run(flow_input)
\end{lstlisting}
\end{figure}

\section{Application Experiences}
\label{sec:app-experiences}

We use five instrument+computation applications to illustrate how the patterns and technologies described in preceding sections can be realized and applied in practice.
These applications link a number of scientific instruments and computing facilities, including Advanced Photon Source (APS)
and Stanford Synchrotron Radiation Lightsource (SSRL)
beamlines and the Argonne Leadership Computing Facility (ALCF)~\cite{riley20192019}.
Each example is implemented by using the Gladier toolkit to define, configure, and manage one or more flows. 
For each, we provide pointers in the Supplementary Information to the source code for both the full application and a simplified implementation that can be run on a personal computer. 

In each of the cases presented here, scientists had previously employed manual and ad hoc methods to implement similar, although typically simpler, behaviors: for example, by capturing data locally, transferring data via portable media to a cluster, and manually running analysis codes. After being introduced to Gladier tools, they implemented, with varying degrees of assistance from Gladier developers, the flows described in the following.

\begin{figure*}
    \centering
    \includegraphics[width=\textwidth,trim=5mm 0 5mm 0,clip]{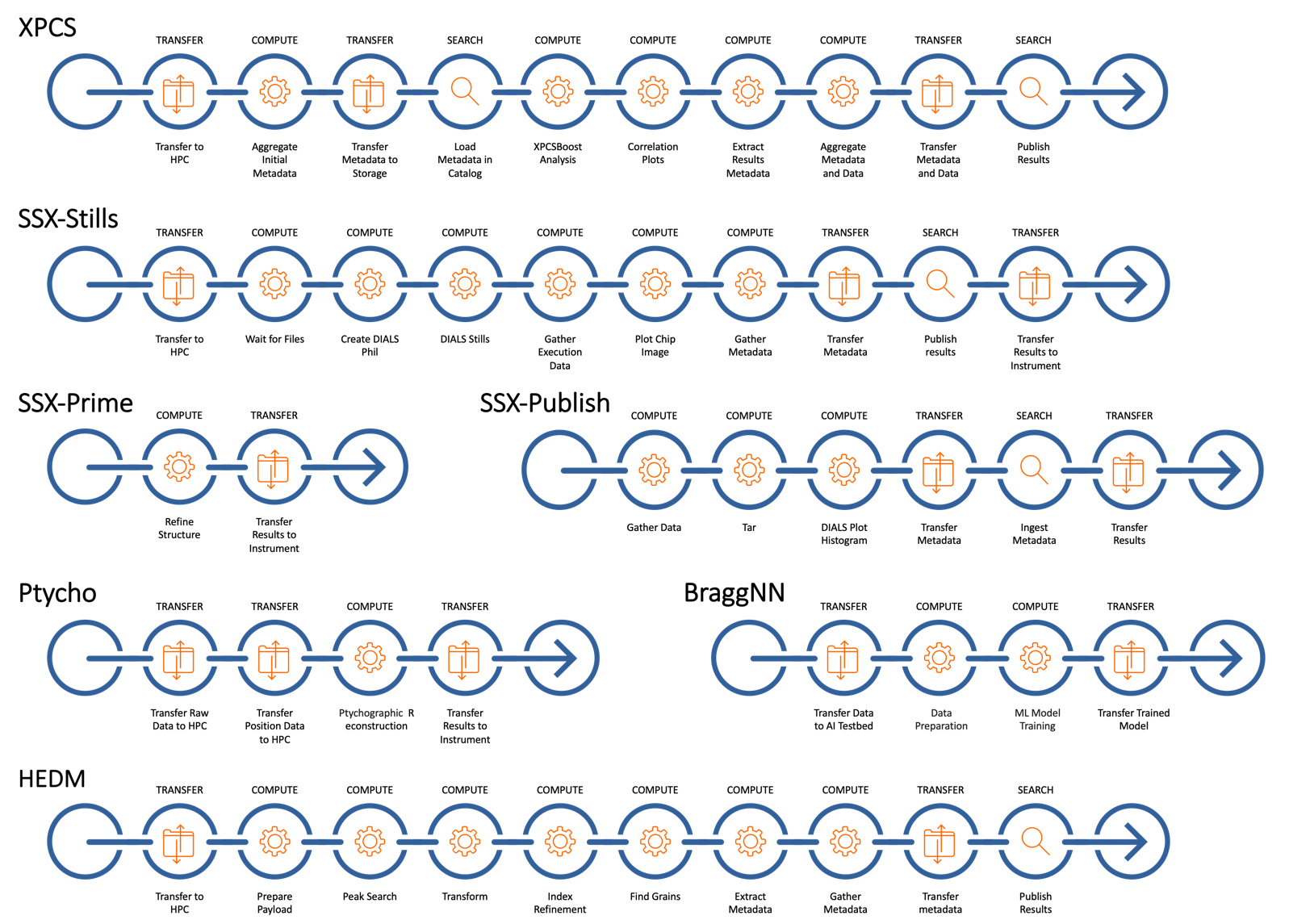}
    \caption{Wireframe depictions of the flows presented in the paper: 
    an x-ray photon correlation spectroscropy processing flow, \textbf{XPCS};
    three serial synchrotron crystallography flows, \textbf{SSX-Stills}, \textbf{SSX-Prime}, and \textbf{SSX-Publish};
    a ptychography image reconstruction flow, \textbf{Ptycho}; 
    a training flow for a neural network function approximator, \textbf{BraggNN}; and
    a high-energy diffraction microscropy far field reconstruction flow, \textbf{HEDM}.
    Text above each circle names the action;
    text below describes its application in the flow.
    }
    \label{fig:all-flows}
\end{figure*}


\subsection{X-Ray Photon Correlation Spectroscopy (XPCS)}

This experimental technique is used at synchrotron light sources to study materials dynamics at the mesoscale/nanoscale by identifying correlations in time series of area detector images~\cite{shpyrko2014x,lehmkuhler2021femtoseconds}.
Current detectors acquire megapixel frames at
up to 2~kHz at 16-bit depth and 50~kHz at 2-bit depth ($\sim$4~GB/s); next-generation detectors are expected to generate 10s of GB/s or more~\cite{perakis2020towards,zhang202120}.
Computing correlations at these rates requires powerful computing, both to process large quantities of data and to enable rapid response for experiment feedback.


We describe a flow developed to automate the collection, reduction, and publication of XPCS data at the APS 8-ID beamline.
Each experiment can produce 100,000s of images, with precise rate and image size controlled by the scientist.
During image acquisition, the instrument's experiment management system typically creates a data file for every 20,000 images ($\sim$2.4~GB); 
to enable use of the automation services described in this paper, it is configured to trigger a flow each time such a file is created.

The flow, illustrated in \autoref{fig:all-flows}, comprises 10 steps:
\begin{inparaenum}[(1)]
\item 
    copy the experiment data file to a compute facility
    ({\sc{Transfer}});
\item
    extract metadata, such as data acquisition parameters and processing instructions, from the experiment data file ({\sc{Compute}});
\item 
    copy these metadata to persistent storage ({\sc{Transfer}});
\item 
    load metadata into a Globus Search catalog, providing visibility into the data that are being processed and the software version and input arguments to be used during subsequent processing steps ({\sc{Search}});
\item 
    run the XPCS Boost correlation analysis function, a matrix-heavy operation that is best run on a GPU ({\sc{Compute}});
\item 
    run a plotting function to create correlation plots and compact images for display in the portal ({\sc{Compute}});  
\item 
    extract metadata from correlation plots ({\sc{Compute}});
\item 
    aggregate the correlation plots, new metadata, execution logs, and compact images for publication ({\sc{Compute}});
\item 
    transfer the aggregated data and metadata to persistent storage ({\sc{Transfer}}); and
\item 
    add the aggregated metadata and associated data references to the catalog entry created in step~4, thus allowing the scientist to verify quality and also making data available for future uses ({\sc{Search}}).
\end{inparaenum}

Before using this flow it must be defined and registered with the Globus Flows service, and any 
tools and infrastructure used by the flow must be installed and configured if not already in place.
We describe these steps in some detail in Section~\ref{sec:SI-xpcs} so as to illustrate the process by which a new flow is configured, deployed, and operated. 
A similar process is required for each of the other applications described in this section.


We note that while all computational steps (2, 5-9) can run on general-purpose CPUs, step 6, XPCS Boost analysis, benefits from use of GPUs and thus the flow is typically configured to access a funcX endpoint associated with a GPU resource.
Using GPUs, the flow can process a dataset and produce visualizations to the scientist in about 240 seconds (see \autoref{tab:overhead}), and in around 50 seconds with dedicated resources.

\subsection{Serial Synchrotron Crystallography}

Serial synchrotron crystallography (SSX) is a technique in which a bright synchrotron beam and specialized apparatus are used to collect diffraction data from many crystals, at rates of 10,000s of images per hour~\cite{diederichs2017serial}.
It can collect diffraction data from samples at room temperature and produce higher quality data than conventional crystallography due to reduced radiation damage~\cite{nam2022serial}. 

We describe here methods used to process SSX data at APS Sector 19.
A typical experiment generates around 40,000 1475$\times$1255 16-bit pixel images per sample, with tens of samples processed during a beamtime.
While the detector is capable of operating at 100Hz, for a data rate of 370 MB/s, the experiment is flux limited and is typically performed at roughly 10Hz, or 37 MB/s.
As images are produced, they are processed (in batches) with the  
DIALS package to identify crystal lattices, or \textit{hits}, in each image.
As hits are accumulated, they are processed with the 
post-refinement and merging (PRIME) package~\cite{uervirojnangkoorn2015enabling} to solve the crystal structure.
DIALS and PRIME outputs are published to an SSX repository and cataloged for subsequent use.


These activities are implemented by three distinct flows.
The first, 
\textbf{SSX-Stills}, transfers a batch of acquired images to a computing facility and uses the DIALS Stills package to perform quality analysis on each image and identify those that contain a good quality diffraction (a \textit{hit}). It comprises 10 steps:
\begin{inparaenum}[(1)]
\item transfer image data from the beamline to a computing facility ({\sc{Transfer}});
\item confirm necessary input files are present ({\sc{Compute}});
\item create configuration files for analysis ({\sc{Compute}});
\item perform DIALS Stills processing on each raw image ({\sc{Compute}});
\item extract metadata from files regarding hits ({\sc{Compute}});
\item generate visualizations showing the sample and hit location ({\sc{Compute}});
\item gather metadata and visualizations for publication ({\sc{Compute}});
\item transfer metadata and visualizations for publication ({\sc{Transfer}});
\item ingest results, metadata, and visualizations to an SSX Globus Search catalog ({\sc{Search}}); and
\item transfer the results back to the beamline ({\sc{Transfer}}).
\end{inparaenum}

The \textbf{SSX-Prime} flow uses diffractions from SSX-Stills to solve the crystal structure. This flow is run first when at least 1000 hits have been identified, and then again to refine the structure as additional hits become available. It:
\begin{inparaenum}[(1)]
\item performs PRIME analysis to solve the structure ({\sc{Compute}}); and
\item copies the structure back to the beamline ({\sc{Transfer}}).
\end{inparaenum}

The \textbf{SSX-Publish} flow publishes results obtained to date, plus derived data such as histograms, to a repository and catalog. Its six steps are:
\begin{inparaenum}[(1)]
\item gather results, metadata, and visualizations ({\sc{Compute}});
\item create an archive file containing processed data ({\sc{Compute}});
\item create histograms of the analysis ({\sc{Compute}});
\item transfer metadata and results for publication ({\sc{Transfer}});
\item publish results to the SSX repository and catalog ({\sc{Search}}); and
\item transfer results back to the beamline ({\sc{Transfer}}).
\end{inparaenum}

These three flows are initiated by a local agent deployed at the instrument that monitors the creation of files.
In the experiments reported here,
an SSX-Stills flow is triggered for each 512 images and an SSX-Publish flow for each 4096 images;
an SSX-Prime flow is triggered initially when at least 1000 hits have been identified, and then again after each SSX-Stills flow completion. 
This flexibility allows each activity to proceed at an appropriate pace, and permits new flows to be triggered given the result of previous flows, further advancing the automation of the scientific process. 

The result is an indexed, searchable collection of processed images and associated statistics that is updated continuously while an experiment is running. Scientists use this catalog to determine whether sufficient data have been collected for a sample, a second sample is needed to produce suitable statistics, or a sample is not producing sufficient data to warrant continued processing~\cite{wilamowski20212}. 

\subsection{Ptychography}
This coherent diffraction imaging technique can image samples with sub-20 nm resolutions~\cite{maiden2011superresolution}. 
A sample is scanned with overlapping beam positions while corresponding far-field \emph{diffraction patterns}, 2D small-angle scattering patterns containing frequency information about the object, are collected with a pixelated photon counting detector.
Current detectors routinely generate 1030$\times$514 12-bit pixel frames at 3~kHz, for $\sim$20 Gbps~\cite{deng2019velociprobe} and TBs per experiment. 
Next-generation detectors will have readout speeds of more than 100~kHz and increased pixel counts, resulting in multi-PB datasets.

Phase retrieval is applied to ptychography data to recover phase information in reciprocal space.
Typical phase retrieval algorithms are iterative and hence computationally expensive. 
ML-based methods that perform phase retrieval in a non-iterative manner~\cite{PtychoNet,nguyen18deep,cherukara18real} can achieve speedups of 10s~\cite{nguyen18deep} to 1000s~\cite{cherukara18real} times,
opening the door to real-time imaging and thus automated steering of experiments.
However, phase retrieval is highly sensitive to material properties,
and hence the ML model must be retrained for each new material. 

The \textbf{Ptycho} flow performs 2D inversion and phase retrieval on diffraction patterns. It comprises three steps~\cite{bicer2021high}:
\begin{inparaenum}[(1)]
\item transfer data from experimental facility to computing facility ({\sc{Transfer}});
\item process each diffraction pattern to obtain a full image ({\sc{Compute}}); and
\item transfer intermediate results back to experimental facility ({\sc{Transfer}}).
\end{inparaenum}
During a ptychography experiment, hundreds of instances of this flow can be initiated concurrently. Further, this flow can be extended with 3D reconstruction steps and science-specific AI/ML methods: for example, feature segmentation and event or phenomena detection to enable feedback loops for experimental steering.

\subsection{High Energy Diffraction Microscopy}

This non-destructive technique combines imaging and crystallography algorithms to characterize polycrystalline material microstructure in three dimensions (3D) and under various in-situ thermo-mechanical conditions~\cite{bernier2011far,pokharel18hedm}. The technique uses a synchrotron beam to map grains in a polycrystalline aggregate by considering diffraction patterns as a function of rotation angle. It thus requires identification of diffraction ``spots'' for each grain. 
Far-field ($\sim$10~$\mu$m) HEDM, near-field ($\sim$1~$\mu$m) HEDM, and tomography may be combined when studying a material~\cite{bernier2011far}, with, for example, far-field data used to guide near-field measurements.

We present two distinct HEDM applications that implement different approaches to HEDM data analysis.
The first, \textbf{HEDM},
involves flows for collection, analysis, and storage of far-field and near-field data, and for coordination of those activities.
We show in \autoref{fig:all-flows} the first of these flows,
which involves eight steps:
\begin{inparaenum}[(1)]
\item transfer data from experimental facility to computing facility ({\sc{Transfer}});
\item process each raw image using MIDAS~\cite{MIDAS} ({\sc{Compute}});
\item extract metadata from files regarding \textit{hits} (identified crystal diffractions) and generate visualizations showing the sample and hit locations ({\sc{Compute}});
\item process each set of processed images (from step~2) to refine structure ({\sc{Compute}});
\item gather metadata ({\sc{Compute}});
\item transfer metadata to storage facility ({\sc{Transfer}});
\item publish raw data, metadata, and visualizations ({\sc{Search}}); and
\item transfer the results back to the experimental facility ({\sc{Transfer}}).
\end{inparaenum}
A single flow typically moves $\sim$11.5~GB and consumes $\sim$400~sec of compute time in steps~2 and~4.

The MIDAS package used by the HEDM application determines peak positions and shapes by fitting the observed intensities in area detector data to a theoretical peak shape such as pseudo-Voigt.
While the HEDM flow presented allows scientists to harness powerful computing for these computations, the higher data rates at new experimental facilities greatly increase overall computational costs~\cite{li2021machine}.
A promising alternative, explored in our second HEDM application, BraggNN, 
is to train and deploy a neural network approximator to the conventional curve fitting function.
The neural network training can be performed on a powerful data center computer (e.g., conventional cluster or AI accelerator), after which the trained network can be deployed on a lightweight ``edge'' device at the instrument for real-time diffraction peak analysis to power applications such as experiment steering and anomaly detection. 

The \textbf{BraggNN} flow, as shown in \autoref{fig:all-flows}, explores the feasibility of this approach and in particular the relative costs of data transfer, network training, and network deployment. It comprises just four steps~\cite{liu2021bridging,nntrain}:
\begin{inparaenum}[(1)]
    \item 
        copy data from beamline to computing facility ({\sc{Transfer}});
    \item 
        prepare the data for training ({\sc{Compute}}); 
    \item 
        train the BraggNN model ({\sc{Compute}}); and
    \item 
        copy the trained model back to the beamline ({\sc{Transfer}}).
\end{inparaenum}
In the experiments described below, data are collected at SSRL and transferred to ALCF for training on AI accelerators such as the Cerebras wafer-scale engine~\cite{lauterbach2021path}.
The ease with which Gladier permits re-targeting of compute tasks proved invaluable when selecting an appropriate platform for different neural network architectures.

\section{Application Usage}
\begin{figure*}[h!]
\centering
    \includegraphics[width=0.99\textwidth,trim=0 0mm 0 0,clip]{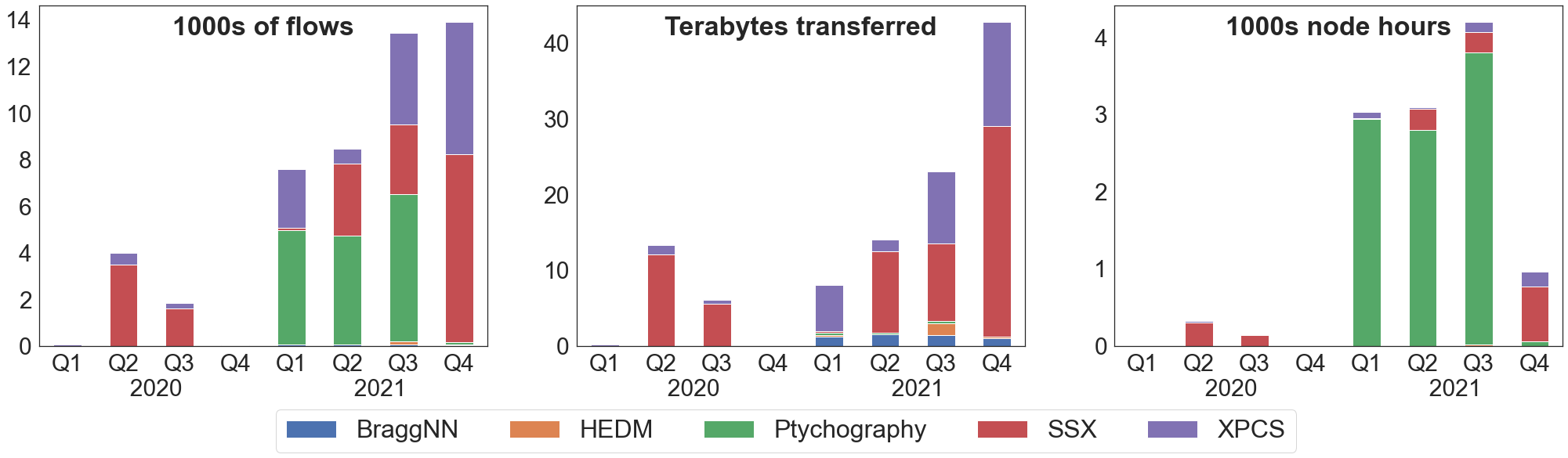}
    \caption{Total flows, data transferred, and compute time used (on 64-core ALCF Theta nodes), per quarter, for the five experiments described in \autoref{sec:app-experiences}.
}
\label{fig:agggb}
\end{figure*}

Scientists have employed the methods and tools described above at APS and ALCF since early 2020 at a cadence that has varied with instrument availability and research priorities, but is generally increasing.
Usage across the five experiments described in this paper, summarized in \autoref{fig:agggb}, encompass 49,367 distinct flow runs that consumed over 11,700 node hours of compute and transferred roughly 108~TB.
The variation in usage across experiments and over time is primarily due to the sporadic nature of experiments at large-scale facilities. There are periods of downtime in which few, or no, experiments are run.  We see a general increase over time in the number of flows run and the amount of data transferred. 
The decrease in compute time in Q4 2021 is due to the fact that the compute-intensive ptychography experiment was not running during this period.
Several experiments are deploying more ambitious and expensive computational methods now that the feasibility of on-demand computing has been established.


\begin{figure}[h!]
\centering
\includegraphics[width=\columnwidth,trim=10mm 10mm 10mm 10mm,clip]{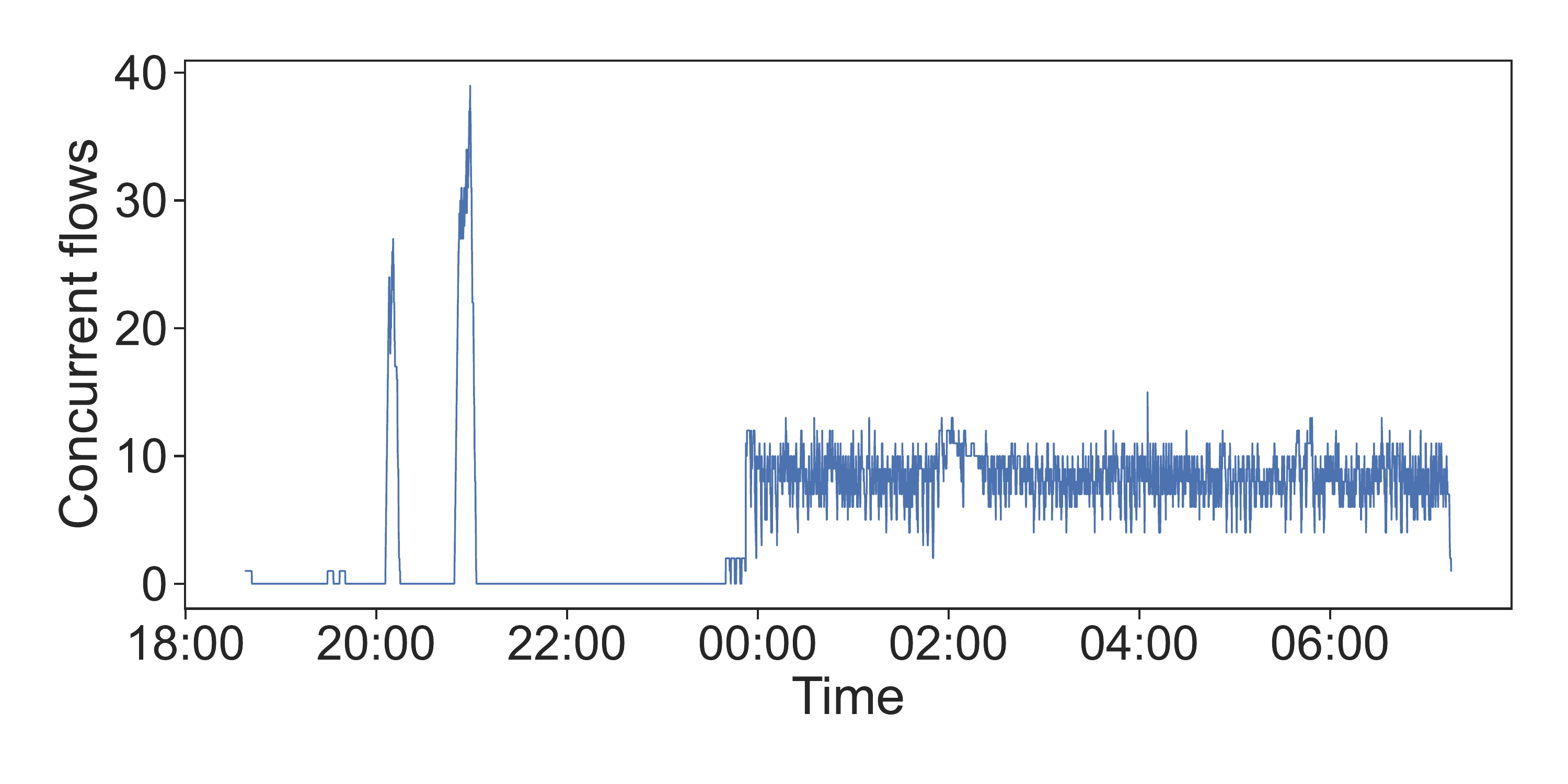}
\caption{The number of concurrent XPCS flows over a roughly 12-hour period,
March 10-11, 2022.
The initial peaks are burst tests before beginning the experiment; by 00:00, a constant stream of data from the beamline is processed.
\label{fig:concurrentflows}}
\end{figure}

We explore in \autoref{fig:concurrentflows} the ability for flows
to keep pace with data acquisition rates. Specifically, we show a twelve hour period in which XPCS flows are executed during an experiment session. 
During a preparatory period of roughly four hours, the scientists run occasional bursts of flows to calibrate equipment and ensure that the analysis pipeline is operational. Here we see up to 39 instances of the XPCS flow executing concurrently, each with the eleven steps shown in \figurename~\ref{fig:all-flows}.
The subsequent eight hours of the experiment, represents steady-state processing in which flows are executed as the result of data acquisition. We see here that approximately 10 flows execute concurrently throughout the experiment, showing that the flows meet the required data acquisition rate of one file per minute. The additional flows represent out-of-band reprocessing tasks executed by the scientists. 

\begin{figure}[h!]
\centering
    \includegraphics[width=\columnwidth]{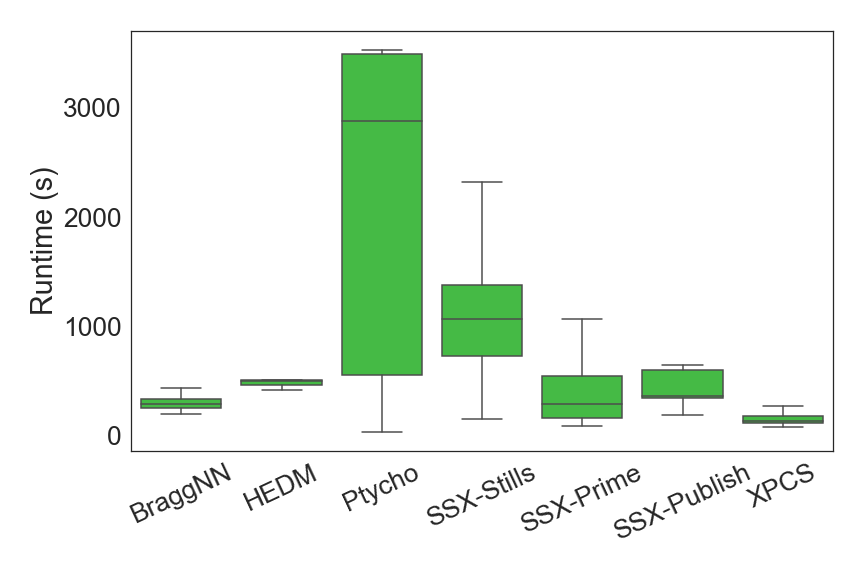}
    \caption{Distribution of runtimes for the seven flows discussed in the text. Box plots show upper and lower quartiles, with whiskers to 1.5$\times$ the interquartile range.
\label{fig:flow_runtimes}}
\end{figure}

We compare the runtime of each flow in \autoref{fig:flow_runtimes}. Here we see mean and quartiles for the more than 2600 flow runs.
We see that the Ptycho flow has significantly longer execution times and also higher variance in execution time (25th to 75th percentile is approximately 2000s) than other flows. This variance is primarily due to unpredictable compute cluster queue delays, as these flows were run without dedicated reservations. Importantly, flows complete reliably despite such delays.

\begin{figure}
\centering
    \includegraphics[width=\columnwidth,trim=10mm 10mm 10mm 10mm,clip]{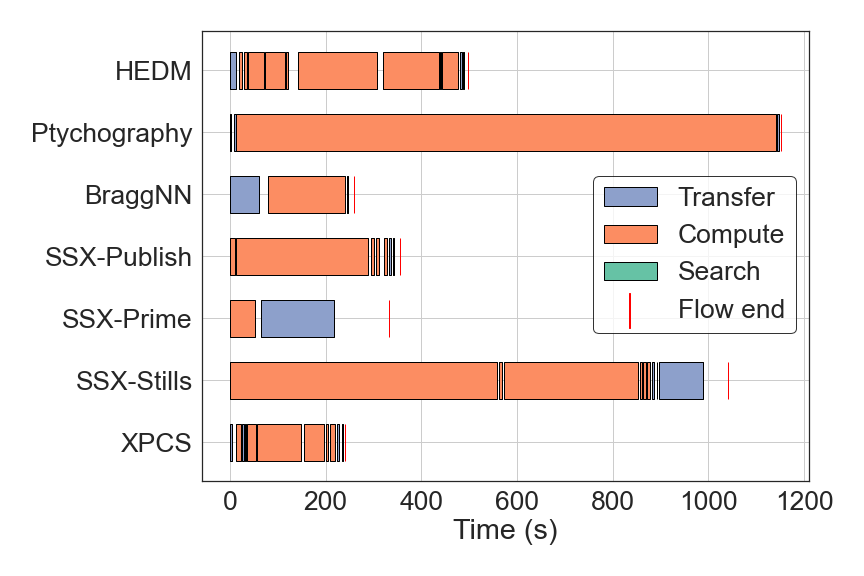}
    \caption{For the instance of each flow with median runtime, a timeline for its constituent actions.
    The empty spaces between steps correspond to flow orchestration overheads.
    \emph{Note that the Ptycho analysis times are scaled to 50\% (from 2261~s to 1130~s total) so as to better show details in the other flows.}
\label{fig:overhead_breakdown}}
\end{figure}

We show in \autoref{fig:overhead_breakdown} a breakdown of action execution time for a single instance of each flow. We select the instance of that flow with median total runtime,
and show the time spent executing each action as measured by the respective action provider. We illustrate overhead as the difference between the time measured by the action provider to perform the task and the time recorded by the Globus Flows service to complete a step.
Overheads include costs incurred as Globus Flows transitions between steps, invokes action providers to submit a task, and, most significantly, polls for action status (see next paragraph). 
Flow durations ranged from a mean of 31s for XPCS to 3527s for Ptycho. All except SSX-Prime are compute bound. For SSX-Prime and some other flows, the overheads (see \autoref{tab:overhead}) reveal opportunities for optimization (e.g., by improved polling strategies) but none are so high as to hinder experiments.

\setlength{\tabcolsep}{4.4pt}
\begin{table}[b]
    \centering
    \caption{For the instance of each flow with median Runtime, the times taken by its constituent Transfer, Compute, and Search action(s), and the aggregate overhead, both in seconds (OH) and as a percentage of total runtime (\%OH).}
    \label{tab:overhead}
\begin{tabular}{lrrrrrr}
\toprule
Experiment & Runtime & Transfer & Compute & Search & OH & \%OH \\
\midrule
    BraggNN & 259.5 & 64 & 162.1 & 0 & 33.4 & 12.9 \\
    HEDM & 498.2 & 16 & 405.9 & 1 & 75.3 & 15.1 \\
    Ptycho & 2283.3 & 11 & 2259.4 & 0 & 13.0 & 0.6 \\
    SSX-Publish & 355.2 & 3 & 306.2 & 1 & 44.9 & 12.7 \\
    SSX-Prime & 332.6 & 152 & 53.7 & 0 & 126.9 & 38.2 \\
    SSX-Stills & 1041.4 & 97 & 860.0 & 1 & 83.4 & 8.0 \\
    XPCS & 240.0 & 12 & 177.9 & 2 & 48.1 & 20.0 \\
\bottomrule
\end{tabular}
\end{table}



\begin{figure*}
\centering
\includegraphics[width=0.99\textwidth,trim=0 130mm 0 0,clip]{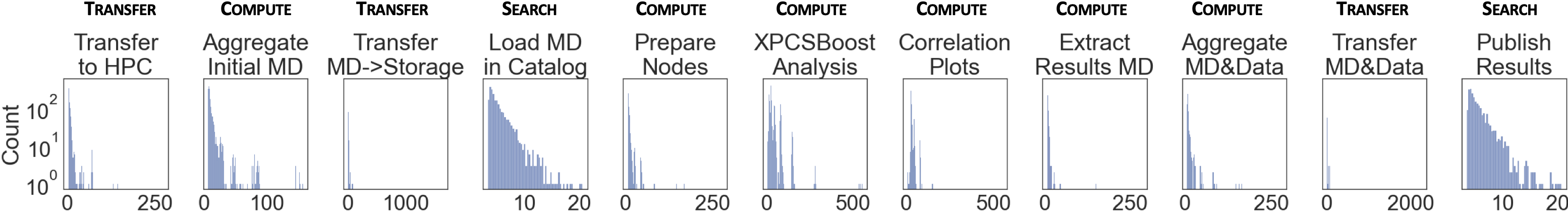}
\\
\includegraphics[width=0.99\textwidth,trim=13mm 0mm 12mm 12mm,clip]{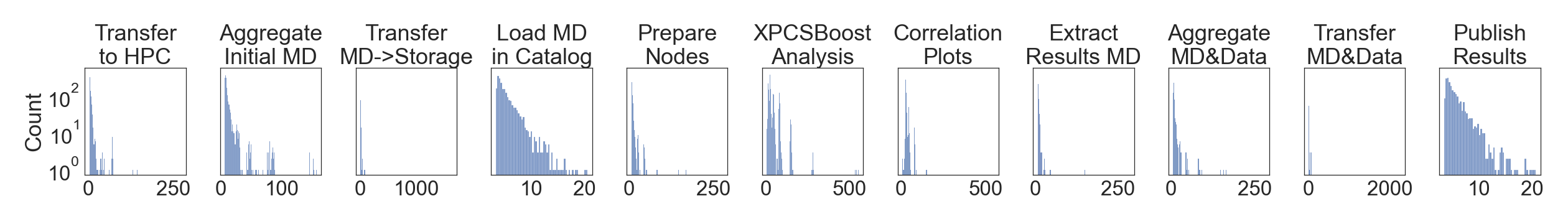}
\\
\includegraphics[width=0.99\textwidth,trim=13mm 0mm 12mm 37mm,clip]{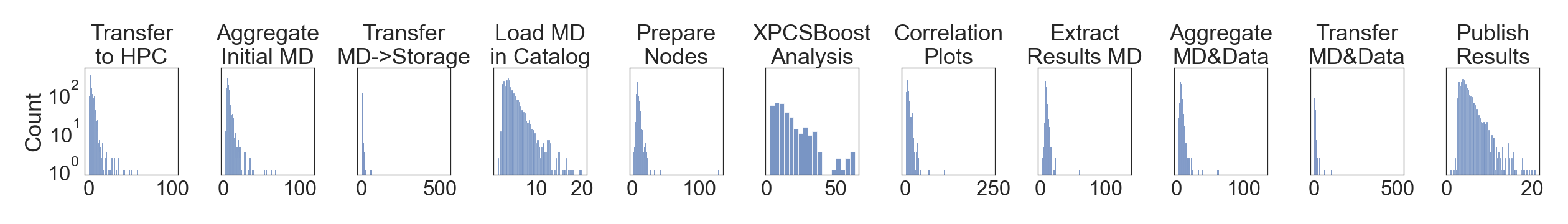}
\caption{Distributions of run time (top) and overhead (bottom), in seconds, for each of the 11 steps in the XPCS flow. MD = metadata.}\label{fig:xpcs_runtimes}
\end{figure*}

\autoref{fig:xpcs_runtimes} drills down on the runtime and overhead of individual steps within the XPCS flow.  The histograms in the top row are of runtimes for each of the flow's 11 steps, over 2197 flow executions;
those in the bottom row are the associated per-step overheads. The varied performance seen in the runtime graphs for transfer and compute actions is expected, as these actions involve functions that may run for minutes and transfers that move gigabytes, and that are subject to compute cluster queue and Globus Transfer limits, respectively. 
The similar distributions seen in the runtime and overhead graphs for the same action are due to the exponential backoff polling interval (starting at 1 second) used by Globus Flows: the longer an action take to execute, the less frequently Globus Flows polls the action to check completion.
(The backoff maximum of 10 minutes is reflected in the maximum overhead of roughly 500 seconds)
The two search actions show more consistent performance (within 20s), although still with outliers.
Roundtrip times to cloud services are not a significant source of overhead for any action.
These results show, again, overheads that are acceptable for these applications, but with opportunities for optimization.

\section{Discussion}
We discuss implications of the patterns and technologies described here for various stakeholders. We base this discussion on our experiences working with the five example applications described in \autoref{sec:app-experiences}, each of which use the patterns and technologies outlined in this paper to meet their science needs.

\subsection{Adopting Patterns}
The patterns presented in this paper can be used to design and implement instrument-linking applications using technologies different than those presented here (i.e., they could be implemented using other components). The patterns illustrate common steps that are necessary for such use cases and outline requirements for related systems. 
Implementations of these patterns present concrete 
examples that can be reused and adapted to other use cases.


\subsection{Adopting Globus and Gladier}
The Gladier toolkit and Globus platform are publicly available and accessible to the research community. Thus, users can define new flows or adapt published flow templates that implement common patterns, including those described here. 
Our platform-based approach means that a user need only ensure that Globus and funcX endpoints are in place before running a flow in a new environment.
At many scientific facilities, required endpoints are already deployed, in which case users need only modify a template to specify endpoints, data locations, and compute functions. In environments where endpoints are not already available, users must first deploy the endpoint software to make their resources accessible---a relatively straightforward task as Globus Transfer endpoint software is distributed in native Linux packages and for MacOS and Windows PCs, while funcX endpoint software can be installed via Python pip (Package Installer for Python).
A happy consequence of these low deployment costs and our use of Python has been considerable diversity in our early adopter community. For example, the flows described in \autoref{sec:app-experiences} were authored by both computer scientists and domain scientists, with little support from our team.

\subsection{Use of a Cloud Platform}\label{sec:platform}

Our use of Globus platform services for IAM, data, flow automation, and computation simplified the realization of the patterns described here. Because Globus operates on a public cloud with publicly accessible APIs and web interfaces, users can readily start, monitor, and manage flows irrespective of where they and their flows are located. They also benefit from the heightened reliability that results from outsourcing the management of multi-step flows spanning distributed resources to a reliable cloud platform with replicated state. 
The cloud-hosted services architecture also makes it easy for users to compose flows in different ways to meet different needs, without the need to apply monolithic software stacks. 

The Globus platform's use of web authentication and authorization standards (e.g., OAuth~2~\cite{oauth2}) provides a rich IAM ecosystem for managing the security of complex flows. This approach allows users and resource owners to manage what actions are performed and by whom, and also supports the complexities of real-world use cases. 
For example, Globus Auth allows for secure integration with external tools (e.g, facility data management systems) by using various OAuth~2 grant types (e.g., for public clients), group-based community accounts for shared computing access, and delegated authorizations for flows to securely invoke external services. 

The ease with which the platform can be extended to edge resources by deploying data and compute agents (Globus collections and funcX endpoints, respectively) is important for use cases that require edge computing. These lightweight and easily installed agents offer crucial capabilities that allow execution of actions on remote and diverse resources. They may be operated by resource owners to support any authorized users, or alternatively deployed by an individual user to process their own requests only.

While the Globus platform provides capabilities needed to implement a broad range of flows, it does not (and cannot) offer \textit{every} capability desired by users. 
Thus, another advantage of the platform model is that we are able to prescribe
a common asynchronous REST API and flexible OAuth-based IAM model such
that others can implement and integrate external actions with the platform. 
This API and IAM model could be used to integrate capabilities provided by other cloud-hosted research platforms,
such as Tapis~\cite{liu2021bridging} and CILogon/CoManage~\cite{basney2014cilogon}. 
Integrating other platforms is dependent on the need for platforms 
to ``trust'' one another so that authorization decisions can be routed to different authorization servers. Adoption of common token formats (e.g., SciTokens~\cite{withers2018scitokens}) would further enable consuming services and agents to validate assertions from different authorization domains.

A potential disadvantage of cloud-based platforms such as Globus is the need for continuous connectivity between research facility and cloud, which introduces a new failure mode and may not be permitted by cybersecurity policies.
We see such concerns declining due to the high availability, reachability, and security of modern clouds, but note that a possible compromise is to use local computers for initial data capture while leveraging the cloud platform for more advanced capabilities.

\subsection{Implications for Computing Facilities}

Rapidly advancing and evolving experimental apparatus and associated computational methods result in growing demands for computing and storage.
The appropriate combination of custom silicon, edge computing, and data center computing likely will evolve over the next decade and beyond; however, it remains natural to turn to large computing facilities (e.g., data centers, clouds) for both capacity and hardware specialization (e.g., accelerators). Such facilities are natural rallying points for data storage and organization coupled with close access to compute resources. These needs are particularly important given the adoption of new computing modalities, such as AI and digital twins~\cite{saracco2019digital,niederer2021scaling}. 

The experiences reported here show the benefits of a platform that permits easy redirection of tasks to different destinations, so that choices can be made based on user preferences and/or institutional policies.
However, enabling such redirection relies on facilities exposing interfaces for remote access to data and computing; IAM infrastructure to enable seamless, yet secure, access to such resources; and methods for enabling access (e.g., to service accounts) without prior direct trust relationships. 


Even simple mechanisms can drive 
innovation. For example, ScienceDMZs~\cite{dart2014science} have enabled unobstructed data flows to/from scientific computing facilities; deployment of user-managed and Globus-accessible storage has allowed scientists to rapidly collaborate using shared data; and support for container technologies has reduced barriers for porting applications between systems~\cite{gerhardt2017shifter}. 
These mechanisms should all be universally adopted by computing facilities to enable instrument+computation flows.

Our work has highlighted other capabilities that could reduce barriers for linking instruments and advanced computing~\cite{uram16}. Flexible, on-demand access to computing capacity is needed to support bursty online workloads. The modest computing demands associated with our five experiments were satisfied at ALCF by a mix of backfill queue, standard queues, and reservations,
but such capabilities may no longer suffice as demands increase.
Some sites operate both specialized queues and dedicated and on-demand clusters~\cite{salim2019balsam,hightower17kubernetes,giannakou2021experiences}, but more flexible scheduling mechanisms are likely needed. In high-demand situations the ability either to transition automatically (through standardized and exposed IAM infrastructure) to other computing facilities, including to the commercial cloud (funcX supports provisioning of cloud instances) without direct intervention from experimental scientists could allow the scientists to stay focused on real-time needs. New facility evaluation metrics are needed that encompass not only utilization but also responsiveness for real-time workloads.

Planning for future computing-enhanced experimental science suffers from inadequate knowledge of future demand and the cost-performance tradeoffs associated with meeting demand in different ways.  
It will be important to establish systematic tracking of resource demand and availability at both experimental and computing facilities.
Also needed is a cohort of staff with expertise in both experimental science and computing to assist with the development, deployment, and executing of flows such as those described here.

\subsection{Implications for Experimental Facilities}

Effective coupling of experiment and computational facilities requires both modern computing infrastructure at experiments and high-quality internal and external network connections; many facilities still have deficiencies in these areas. 
Adoption of the ScienceDMZ architecture~\cite{dart2014science, chard2018modern} 
is important so as to eliminate bottlenecks in network paths. 
Experimental facilities must support deployment of the Globus and funcX software needed to integrate with the cloud-based compute and data fabric described here. This is both a social and technical challenge. Administrators must allow for policies that permit deployment and provide for external connectivity, both to computing facilities and to cloud-hosted platform services. Facilities must provision hardware near instruments so that agents can be deployed close to data sources.
Work is also needed to integrate IAM ecosystems. Many facility users are locked within a single IAM domain. Adoption of federated IAM, such as that provided by Globus Auth, and adopted by a growing number of scientific computing facilities, can integrate diverse IAM domains. By adopting standard mechanisms, facilities can make their identities accessible to modern cloud platforms. 

There are opportunities for yet more sophisticated integration. For example, direct integration of the methods described here with the software tools employed by scientists reduces barriers for use by providing familiar interfaces to automation capabilities. Flows can also be used to control experiments, a practice that will require implementation of common APIs, perhaps aligning with the action provider API, for instruments and other devices. 

Full automation (without human intervention) will require that experiments generate meaningful events that can be used to trigger flow executions~\cite{chard2018high}. In the applications reported here, flows are triggered by mechanisms that monitor co-located file systems to integrate with beamline software. 
Other integrations are possible, such as connecting with instrument control systems like EPICS~\cite{epics}, Bluesky~\cite{allan2019bluesky}, LabView~\cite{kodosky2020labview}, and ROS~\cite{quigley2009ros} that allow for generation of events. 

\subsection{Implications for Scientists}

Higher data acquisition rates, larger datasets, and more complex processing flows mean that scientists must increasingly embrace automation to remain competitive. 
The outsourcing of automation tasks to cloud-hosted platforms, as described here, can simplify this transition by avoiding the need for larger local hardware and software deployments. 
However, scientists must be willing to trust external providers to handle mission-critical functionality.
The growing reliance on cloud-hosted services in our daily lives, coupled with their extreme availability and reliability, helps to expedite this transition.  

Adopting the patterns and methods proposed here requires that scientists decouple traditionally monolithic workflows into series of discrete steps that may be executed separately. This approach can improve understandability and make it easier to substitute implementations for individual steps (e.g., to update an analysis routine) and to execute steps in more preferable locations (e.g., in terms of cost, availability, performance). 

We see increasing use of ML techniques for data analysis and for selecting experiment configurations, samples, and processes, with an increasing focus on completing the feedback loop to enable automated steering of experiments. 
These developments make it yet more important to automate data capture and cataloging so as to provide a clear provenance path when data are used for ML model training.

\subsection{Facilitating FAIR Science}

\begin{figure}
\centering
\includegraphics[width=\columnwidth,trim=0 40mm 0 0,clip]{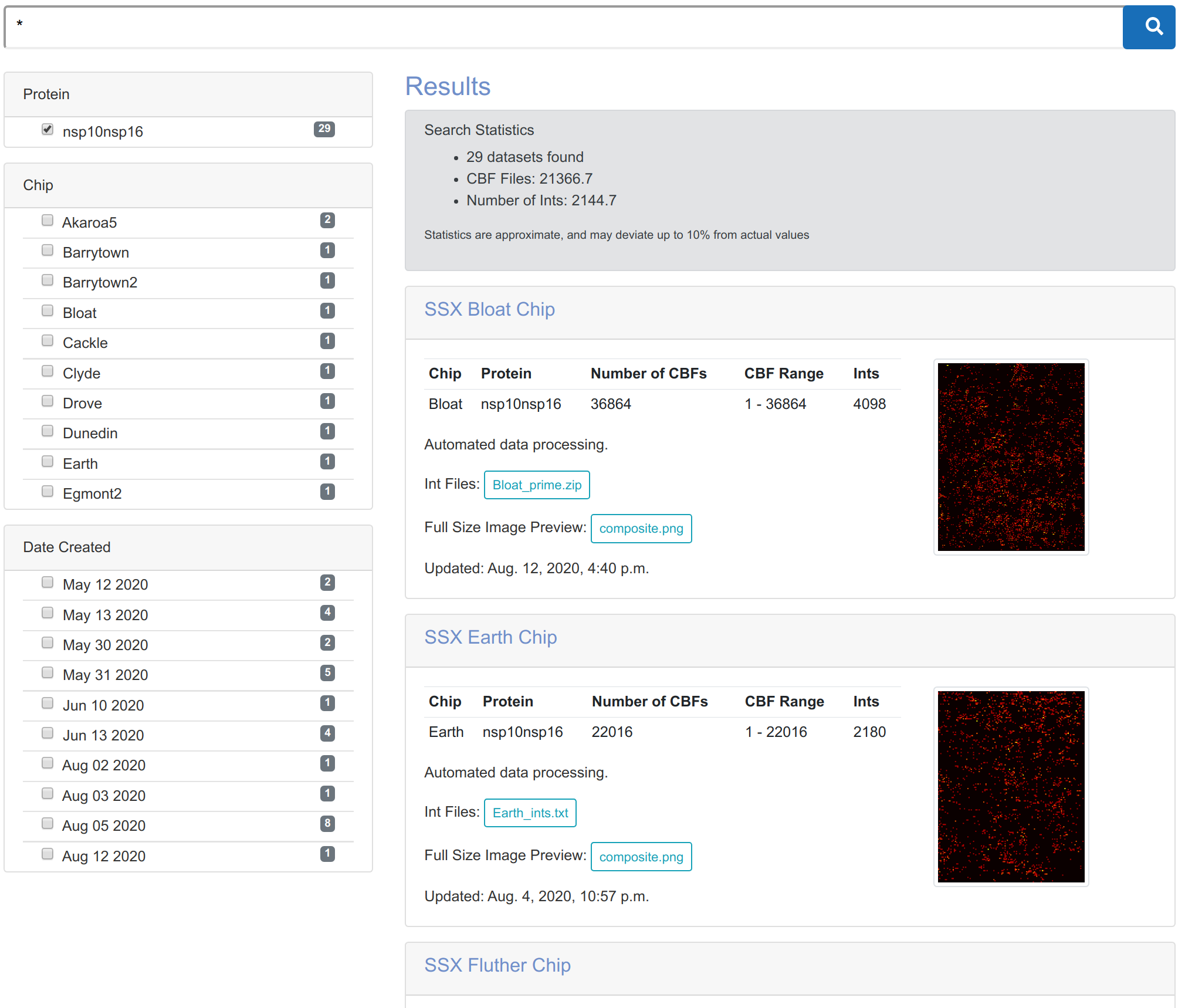}
\caption{SSX data analysis portal. 
Facets on the left allow for selection of different proteins (\textit{nsp10nsp16} is selected here), chips, 
and creation dates. Search results, shown on the right, provide researchers with a quick summary of the experiment and visual representation of the analysis results}\label{fig:kanzus}
\end{figure}

The methods described in this article can contribute to making experimental data findable, accessible, interoperable, and reusable (FAIR)~\cite{wilkinson2016, brinson-fair-comment} by making it easy to integrate data publication into data acquisition and analysis flows. 
In the SSX, HEDM, and XPCS examples presented here, data plus descriptive metadata (expressed in an extensible schema based on that of DataCite~\cite{dcschema})
are published automatically to a Globus Search catalog, with an auto-generated interactive portal: e.g., see \autoref{fig:kanzus}.
These catalogs have been used to index collections containing many terabytes in thousands of files.
Trained models can also be published~\cite{ravi2022fair}.

\section{Related Work}

Specialized data processing systems have been developed in fields such as high energy physics~\cite{bird2011computing} and very long baseline interferometry~\cite{schuh2012vlbi}.
At the Large Hadron Collider, $\sim$1~PB/s data streams are reduced by custom electronics and then 
processed on a distributed computing grid with 100,000s of cores~\cite{bird2011computing}.
More routine
linking of instruments with computers~\cite{johnston1997high,las99siam,goscinski2014multi,toby2015practices} predates the Internet~\cite{dessy1977computer}.
Automation has involved both experiment-specific code~\cite{basu2019automated,khan2018distributed} and orchestration and analysis solutions targeted at specific communities~\cite{benecke2014customizable,gursoy2014tomopy,deslippe2014workflow,talirz2020materials}.
However, none enable specification and reuse of end-to-end flows as here.

Experimental facilities use control systems such as EPICS~\cite{epics} to drive instruments and monitor experiments. Bluesky~\cite{allan2019bluesky} provides Python interfaces for experiment control and data collection~\cite{olds2021optimizing}. These systems can be combined with analysis tools and workflow systems to process data as they are captured. 
Streaming protocols can be used to expedite data movement~\cite{buurlage2019real,scistream}.

The Globus data fabric on which we build here is widely deployed in the US and other countries~\cite{chard2014efficient}.
Other data sharing approaches, varying in scope, maturity, and adoption,
include logistical networking~\cite{beck2000logistical},
Rucio~\cite{barisits2019rucio} and StashCache~\cite{weitzel2019stashcache} in high energy physics, 
ELIXIR~\cite{harrow2021elixir} for the life sciences,
PANdata~\cite{bicarregui2015pandata,pandata} and EXPANDS~\cite{expands} for photon and neutron science, 
iRODS~\cite{xu2017irods},
and the European Open Science Cloud~\cite{eosc}.

The term \textit{scientific workflow} encompasses many technologies~\cite{wainer1996scientific, barker2007scientific, zhao2008scientific, deelman2009workflows}. 
Scientific workflow systems are commonly used to orchestrate many-task computational campaigns~\cite{deelman2015pegasus,wilde2009parallel,goecks2010galaxy} that may execute local programs or submit jobs to data center computers.
Research on workflow scheduling, execution, and related problems has enabled impressive scale and performance within individual systems or across multiple computers under coordinated control~\cite{thain05condor,frey2002condor}.
In contrast, the patterns that are our focus engage many concerns besides orchestration of compute jobs~\cite{stansberry2019datafed}.
We require methods for linking diverse activities and resource types, from computations on computers to experiments on scientific instruments; 
integrating different resource types;
bridging authentication domains;
managing flows that may run for days or even weeks; and 
organizing and arbitrating among collections of flows.
These concerns motivate our decision to build on the cloud-hosted Globus platform~\cite{ananthakrishnan2015globus},
that provides for robust orchestration of diverse activities managed by purpose-specific agents that are already widely deployed.
(The Taverna Web services orchestration platform, while not cloud hosted, had similarities~\cite{oinn2004taverna}.)
The extensibility of the Globus platform allows for the introduction of new non-compute elements into flows and thus into the patterns realized by these flows.

Bridging instruments and distributed computation requires capabilities for reliable and secure remote task submission. 
This challenge motivated Grid computing~\cite{foster2011history,shiers2007worldwide} and the superfacility concept~\cite{enders2020cross}. 
Facilities have developed specialized interfaces for remote job submission~\cite{cholia2010newt,stubbs2021tapis} 
and for managing 
workloads on and across
systems~\cite{thain2005distributed,nickolay2021towards}.
Remote execution has been integrated with Jupyter
notebooks~\cite{Kluyver2016jupyter,parkinson2020interactive,thomas2021interactive}. 
The ability to compute anywhere enables users to leverage specialized computing resources designed for low-cost, distributed, and edge computing~\cite{pordes2007open}. AI systems deployed at experimental facilities support rapid data filtering at the edge~\cite{liu2021bridging}.


Domain-specific data repositories can play a pivotal role in fostering collaboration~\cite{jain2013commentary,de2018tomobank,blaiszik2019data1}. Science gateways~\cite{wilkins2008teragrid,marru2011apache} address data and compute challenges by abstracting underlying resources and providing intuitive analysis interfaces.

The value of federated identity and single sign on as means of streamlining access to scientific resources is broadly recognized~\cite{welch2011roadmap,broeder2012federated,linden2018common,umbrella}, although not yet universally adopted.
Globus Auth complements such initiatives by using OAuth tokens~\cite{oauth2} to delegate to third parties (e.g., a funcX server) the right to perform certain tasks, such as transferring data and running functions, on a user's behalf.
Delegation methods have been developed previously~\cite{gasser1990architecture,foster1998security,welch2004x}.

\section{Summary}
Maximizing the value obtained from new instruments requires tight coupling with heterogeneous and large-scale computing facilities, and new online computing methods to automate data collection, processing, and dissemination.
We have reported on our experiences working with five groups of instrument 
scientists, first to understand their current and future computing challenges and second to automate various of their research \textit{flows}.
We described an automation approach that leverages Globus platform services to enable construction of flows by composing modular components that execute programs, transfer files, publish data to catalogs, manage data permissions, and generate persistent identifiers, among other tasks.
Importantly given dynamic resource availability, our approach achieves a
separation of concerns between \textit{what} actions are applied in each flow and \textit{where} those actions are performed. 
We also described Gladier, a Python toolkit that
abstracts registration of funcX functions, flow authoring, and flow execution with specific input arguments, and simplifies the coupling of such flows to experiments.

The five experiments discussed here vary significantly in their data rates, flow and action runtimes, use of heterogeneous resources, and geographically distributed execution. 
We provide quantitative evaluations of those differences, and demonstrate that our methods can in each case support the robust, scalable, and performant execution required for production use, with overheads that they are acceptable even for complex and long-running flows.

This work represents a first step towards identifying, and capturing in reusable forms, a broad collection of patterns for processing data from scientific instruments---patterns that range from online data processing to machine learning training and data cataloging.
We believe that understanding these patterns and the methods and resources required to support their execution will have important implications for a range of stakeholders, from individual scientists to compute facilities, experimental facilities, and cloud-based research platforms. 

\section*{Supplemental Information Description}

The Supplemental Information provides:
illustrations of the Globus Flows user interface (\ref{sec:flowsUI});
a description of the steps involved in deploying simplified versions of the five applications described in the paper (\ref{sec:SI-xpcs}); and 
details on data and code (\ref{sec:SIflows} and \ref{sec:SIsoftware}).

\section*{Acknowledgements}
This work was supported in part by NSF grants OAC-1835890 and OAC-2004894; award 70NANB14H012 from the U.S. Department of Commerce, National Institute of Standards and Technology as part of the Center for Hierarchical Material Design (CHiMaD); and by the U.S.\ Department of Energy under Contract DE-AC02-06CH11357, including by the Office of Advanced Scientific Computing Research's Braid project. 
We are grateful to staff at the Advanced Photon Source, Argonne Leadership Computing Facility, University of Chicago Globus group, and Stanford Synchrotron Radiation Lightsource for assistance with this work. 

\subsection*{Author Contributions}



RV: Software, Investigation, Writing -- review \& editing;
RC: Conceptualization, Software, Investigation, Writing -- review \& editing;
NDS: Software, Investigation;
BB: Conceptualization, Project administration, Methodology, Writing -- review \& editing;
JP: Software, Writing -- review \& editing;
TB: Investigation;
AL: Investigation;
ZL: Investigation;
MEP: Conceptualization, Writing -- review \& editing;
SN: Investigation;
NS: Conceptualization, Project administration;
KC: Conceptualization, Writing -- original draft, review \& editing;
ITF: Conceptualization, Methodology, Writing -- original draft, review \& editing; Project administration.



\section*{Materials availability}

This study did not generate new unique reagents.

\section*{Data and code availability}

We provide in Sections~\ref{sec:SIflows} and \ref{sec:SIsoftware} pointers to the data and code required to reproduce the results presented in this paper.

\bibliographystyle{unsrt}
\bibliography{refs}

\newpage
\pagestyle{empty}

\appendix
\renewcommand{\thesection}{SI-\arabic{section}}    
\noindent 
\begin{center}
{\large{\textbf{Linking Scientific Instruments and Computation:\\
Patterns, Technologies, Experiences}\\
\vspace{1ex}
\emph{Supplementary Information}}}
\end{center}

\vspace{1ex}

\renewcommand{\thefigure}{SI-\arabic{figure}}
\setcounter{figure}{0}
\renewcommand{\thetable}{SI-\arabic{table}}
\setcounter{table}{0}

\noindent
We provide here supplementary information relating to the paper, \emph{Linking Scientific Instruments and Computation:
Patterns, Technologies, Experiences}.
See the paper for many more details.

\section{Globus Flows web interface}\label{sec:flowsUI}

\begin{figure}
    \centering
    \begin{subfigure}[b]{0.48\textwidth}
        \setlength{\fboxsep}{0pt}\fbox{\includegraphics[width=\textwidth]{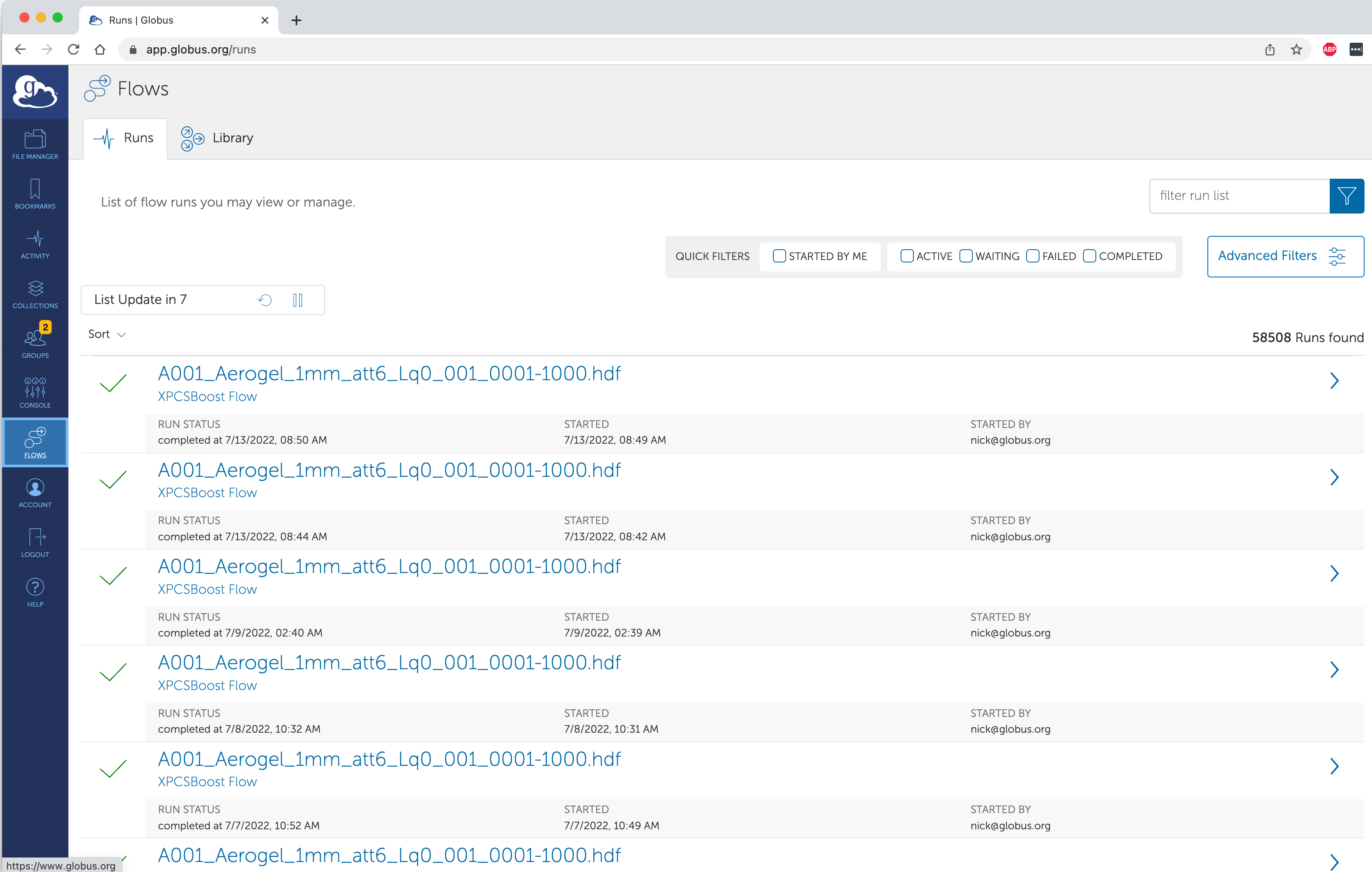}}
        \caption{The \textbf{Runs} tab in the Flows interface lists runs that I can view or manage. The \textbf{Library} tab lists flows that I can run.}
        \label{fig:XPCS_flows_a}
    \end{subfigure}
    \hspace{2mm}
    \begin{subfigure}[b]{0.48\textwidth}
        \setlength{\fboxsep}{0pt}\fbox{\includegraphics[width=\textwidth]{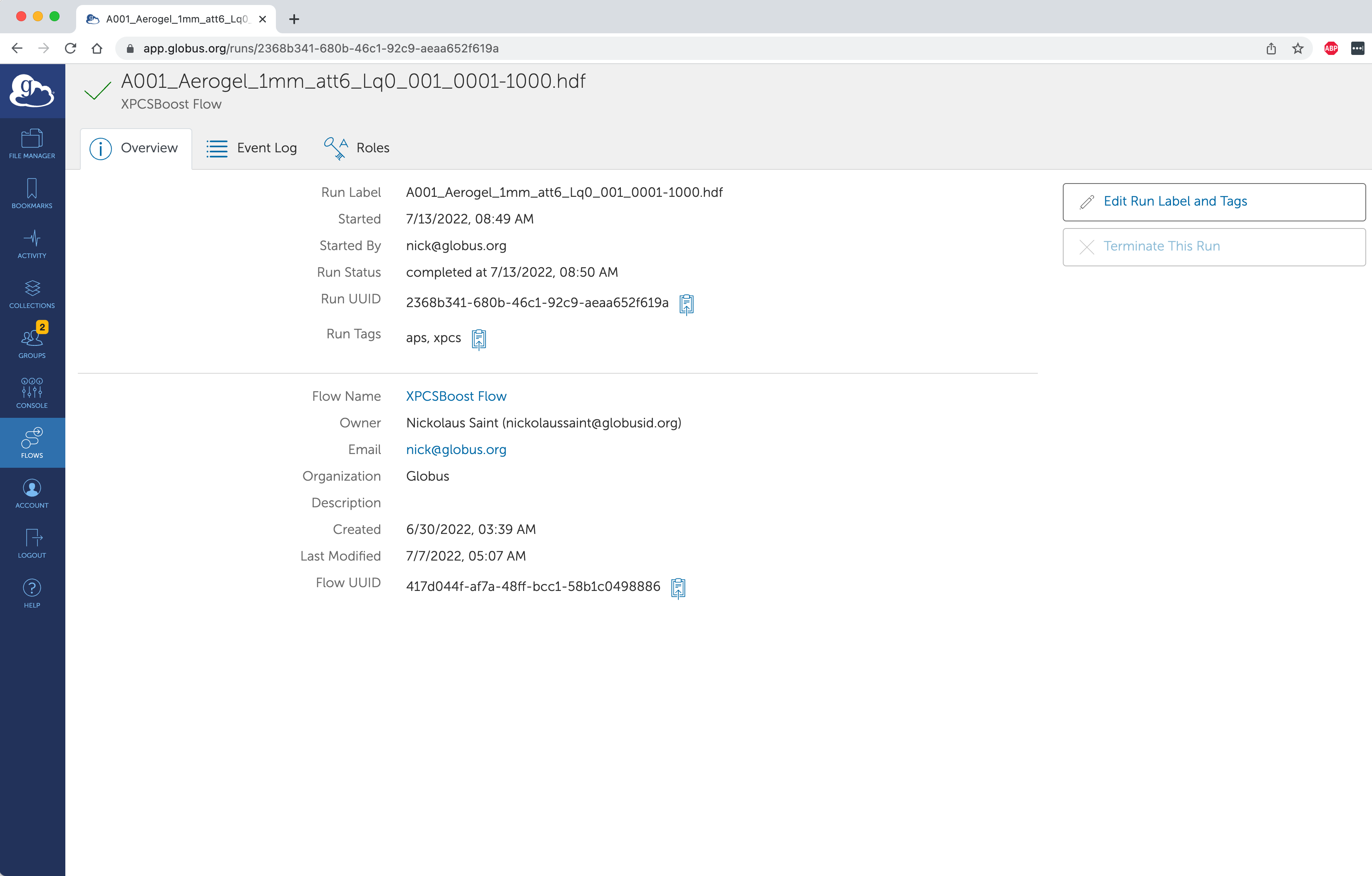}}
        \caption{Selecting a run in \autoref{fig:XPCS_flows_a} gives this status summary, with information on the run (above) and the flow that was run (below).}
        \label{fig:XPCS_flows_b}
    \end{subfigure}
    
    \vspace{2mm}
    
    \begin{subfigure}[b]{0.48\textwidth}
        \setlength{\fboxsep}{0pt}\fbox{\includegraphics[width=\textwidth]{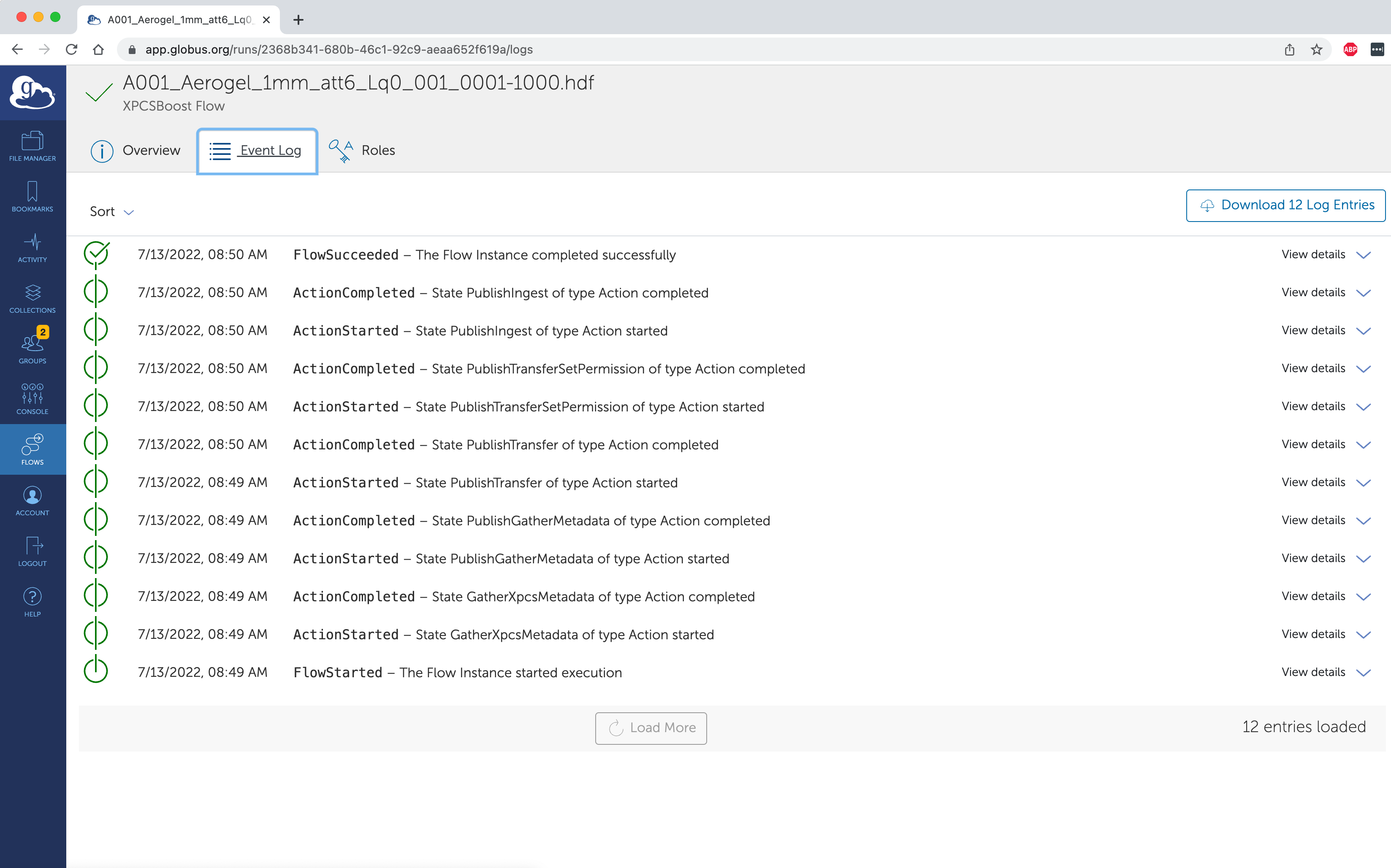}}
        \caption{Selecting the \textbf{Events} tab in \autoref{fig:XPCS_flows_b} gives this list of events during the run. We see that all completed successfully.}
        \label{fig:XPCS_flows_c}
    \end{subfigure}
    \hspace{2mm}
    \begin{subfigure}[b]{0.48\textwidth}
        \setlength{\fboxsep}{0pt}\fbox{\includegraphics[width=\textwidth]{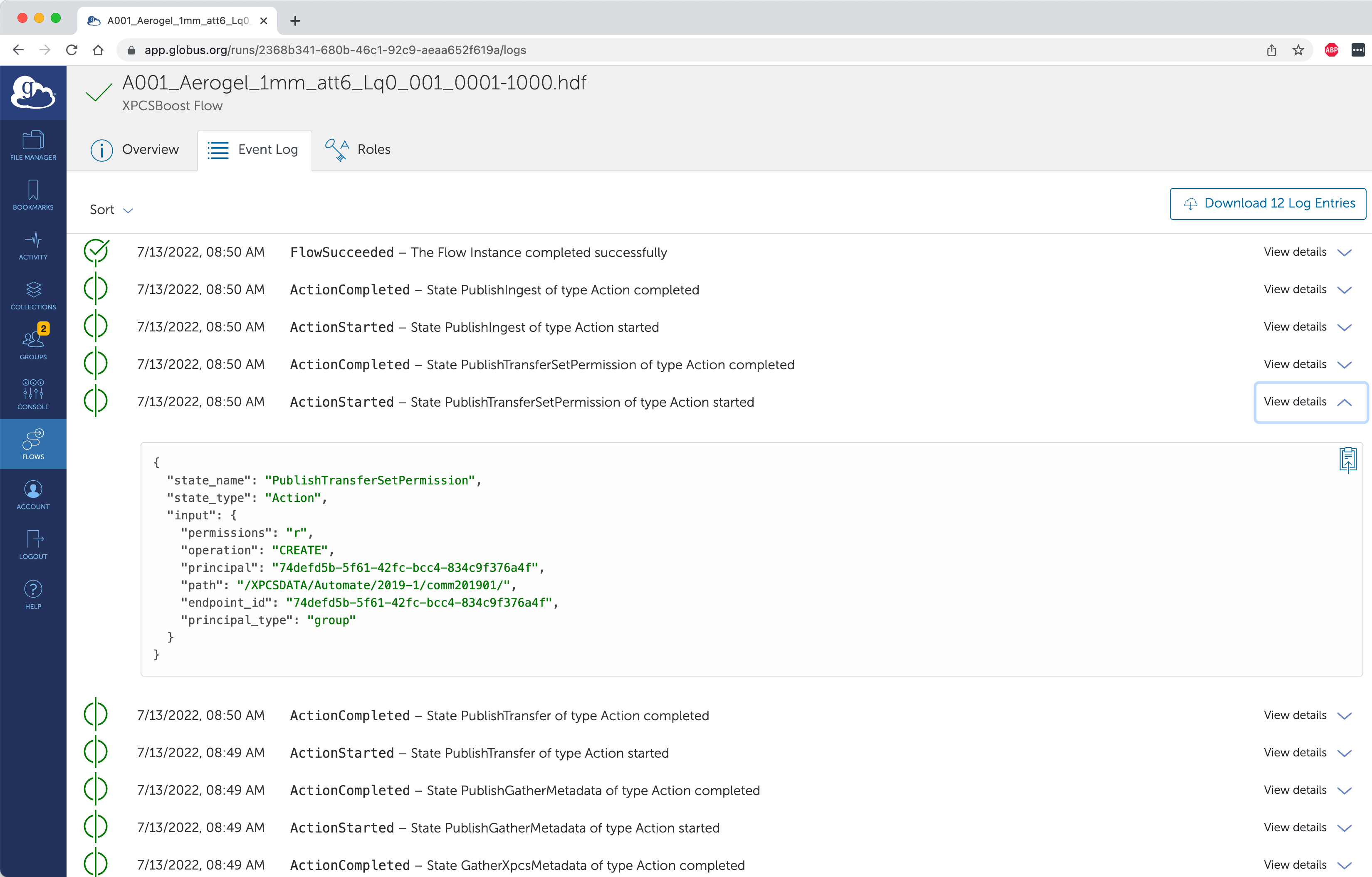}}
        \caption{Selecting a single event in \autoref{fig:XPCS_flows_c} provides additional information about the associated action: \texttt{PublishTransferSetPermission}.}
        \label{fig:XPCS_flows_d}
    \end{subfigure}
    
    \vspace{2mm}
    
    \begin{subfigure}[b]{0.48\textwidth}
        \setlength{\fboxsep}{0pt}\fbox{\includegraphics[width=\textwidth]{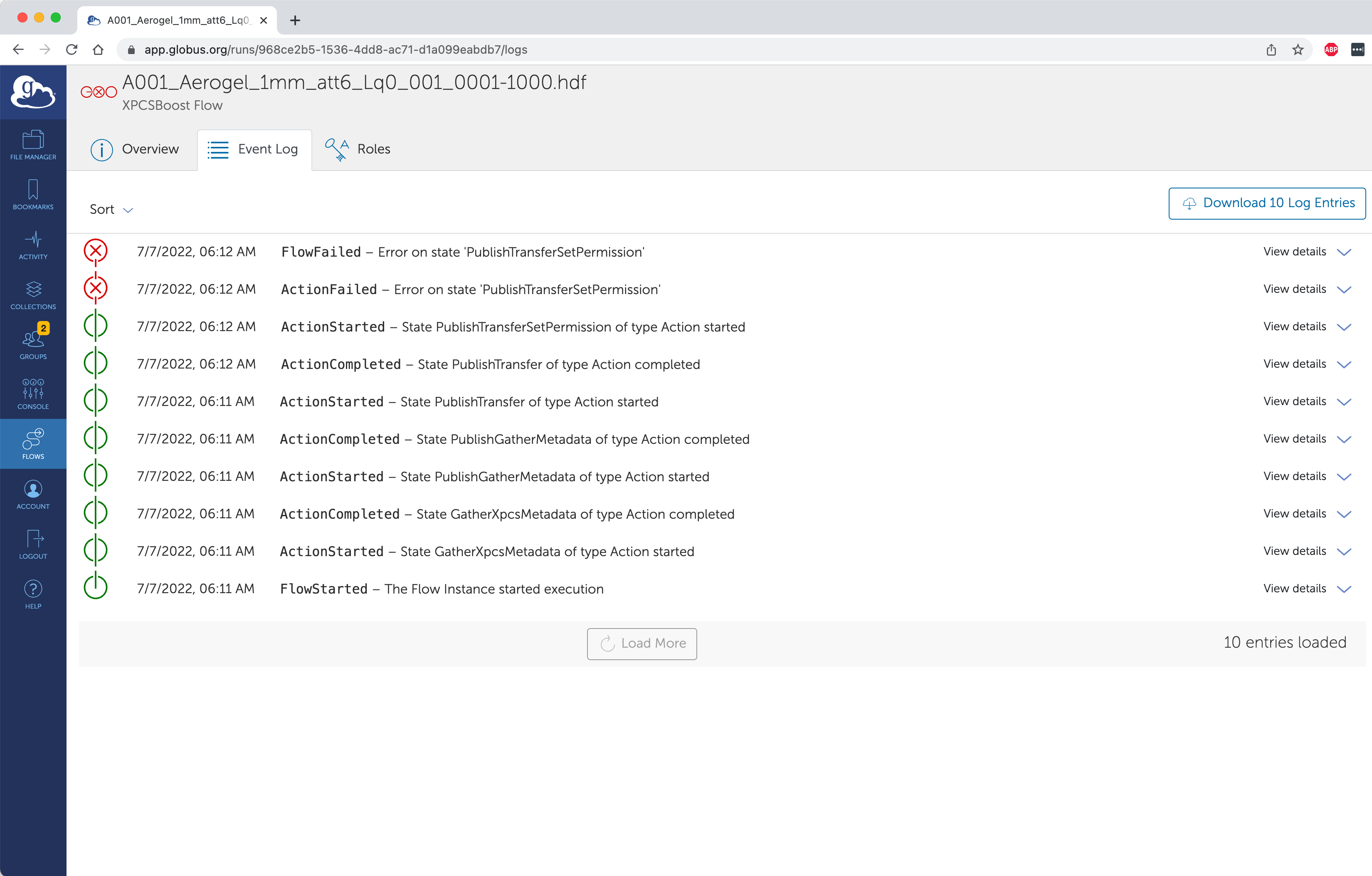}}
        \caption{The events list for an alternative, unsuccessful run of the same flow indicates that an \texttt{PublishTransferSetPermission} action failed.}
        \label{fig:XPCS_flows_e}
    \end{subfigure}
    \hspace{2mm}
    \begin{subfigure}[b]{0.48\textwidth}
        \setlength{\fboxsep}{0pt}\fbox{\includegraphics[width=\textwidth]{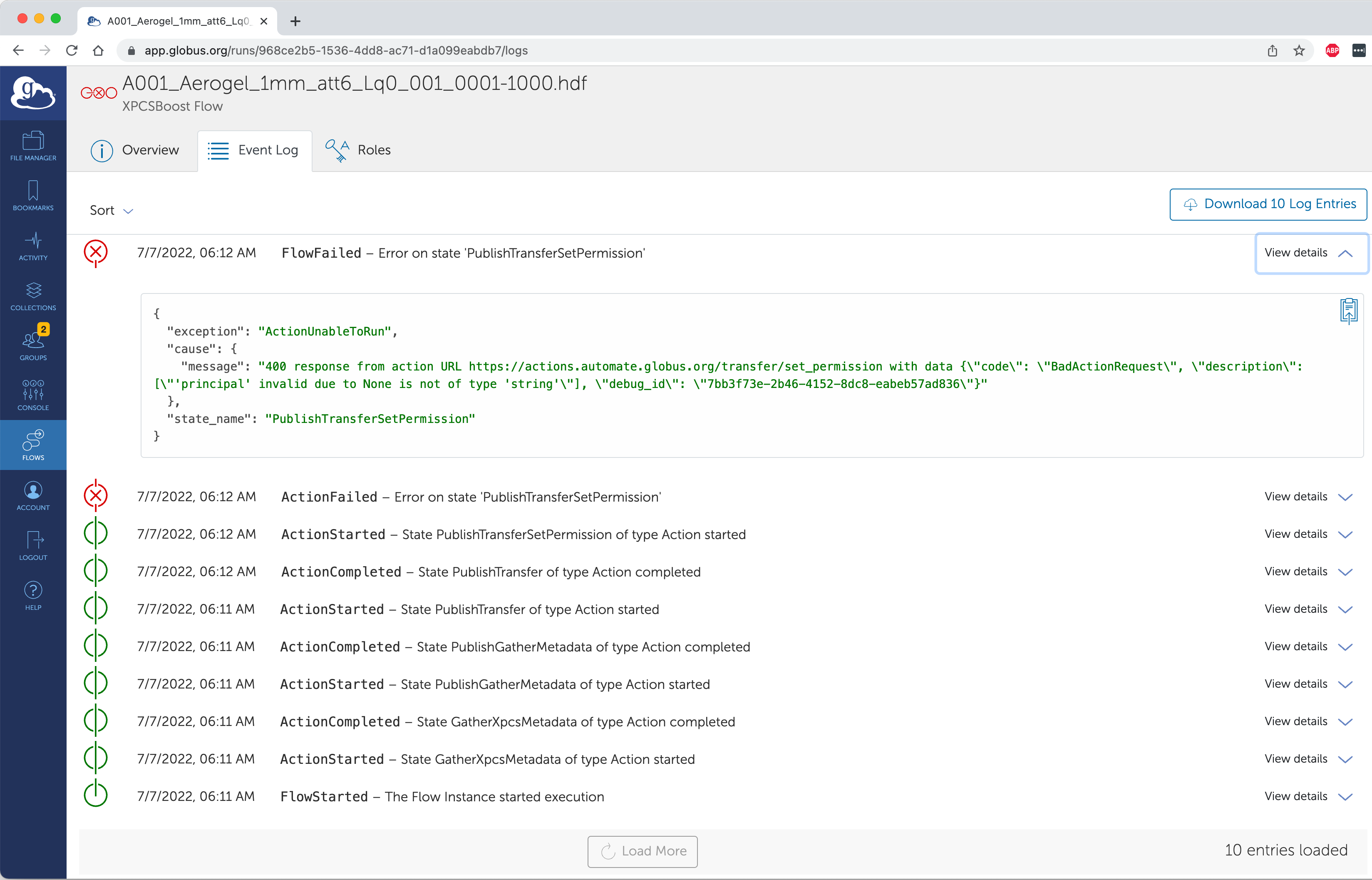}}
        \caption{Drilling down on the erroneous event in \autoref{fig:XPCS_flows_e} reveals (arguably opaque) information about the error: an invalid credential.}
        \label{fig:XPCS_flows_f}
    \end{subfigure}  
    
    \caption{We use the example of an XPCS flow to illustrate how the Globus web interface enables tracking of flow progress and diagnosing of errors.
    }
    \label{fig:XPCS_flows}
\end{figure}

\noindent
Scientists need to be able not only to run flows but to detect, diagnose, and correct errors that may occur when a flow is executing.
The Globus Flows service that we use to run flows provides such capabilities, as we illustrate in \autoref{fig:XPCS_flows},
in which we 
(a) list recent runs,
(b) inspect a summary of a run, and
(c, d) list all actions involved in that run; and
(e, f) examine actions performed in an unsuccessful run.
Other displays, not shown here, allow for examination of flow definitions and input schema. 

\section{Running simplified versions of our five example applications}\label{sec:SI-xpcs}


\noindent
The five applications described in this paper, for which we provide links to source code in \ref{sec:SIflows}, have been developed to process big data streams from real light source instruments. To facilitate exploration, we also provide simple versions of each application that can be configured to run on a personal computer.\footnote{\label{repo:xpcs}\url{https://github.com/globus-gladier/gladier-patterns-examples-2022}} For simplicity, these simplified applications do not deal with publishing flow products to a Globus Search catalog, and they do not have an associated portal.

We first use a simplified version of the XPCS application described in the body of the paper to illustrate how the Gladier toolkit is used to implement a flow, and the process by which a flow is configured and run.
Then, we provide brief notes on each of the other simplified applications

\subsection{The simplified XPCS application}

\noindent
The simplified XPCS application, \texttt{simple\_xpcs\_client.py},\footnote{\label{fn:xpcs_client}\url{https://github.com/globus-gladier/gladier-patterns-examples-2022/blob/main/simple_xpcs_client.py}} involves just three steps, as follows:

\begin{center}
\includegraphics[width=0.5\textwidth]{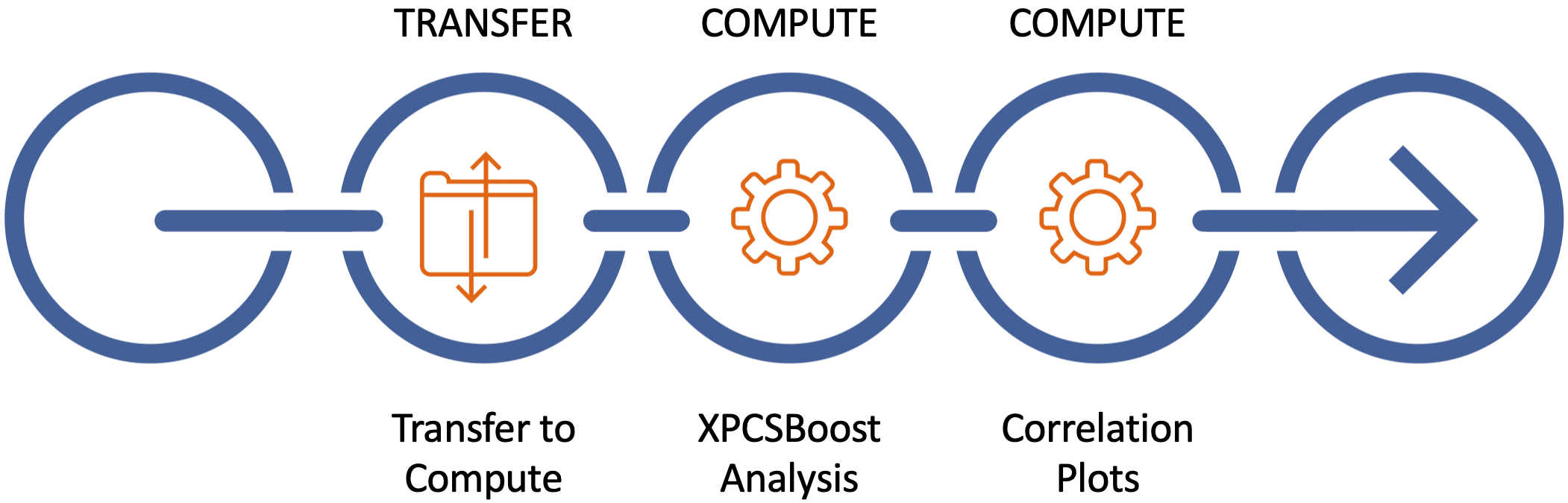}
\end{center}

Consider first the following lines of \texttt{simple\_xpcs\_client.py}:

\begin{lstlisting}[language=json]
 4  from gladier import GladierBaseClient, generate_flow_definition
 5  from tools.xpcs_boost_corr import BoostCorr
 6  from tools.xpcs_plot import MakeCorrPlots
 7   
 8   
 9  @generate_flow_definition
10  class XPCSBoost(GladierBaseClient):
11    gladier_tools = [
12       "gladier_tools.globus.transfer.Transfer:FromStorage",
13       BoostCorr,
14       MakeCorrPlots,
15    ]
\end{lstlisting}

Lines 11-15 of this code uses the Gladier toolkit to specify a flow comprising the three tools shown in the figure:
\begin{enumerate}
    \item 
    A {\sc{Transfer}} task to move data from a source storage location to a destination storage location (line 12). In a real deployment, the source will typically be a storage system associated with the scientific instrument and the destination a storage system associated with the data center where the analysis computer(s) are located.
    \item 
    A first {\sc{Compute}} task to run the XPCSBoost Analysis program (line 13; imported, as specified in line 5, from \texttt{tools/xpcs\_boost\_corr.py}\footnote{\url{https://github.com/globus-gladier/gladier-patterns-examples-2022/blob/main/tools/xpcs_boost_corr.py}}).
    \item 
    A second {\sc{Compute}} task to run the Correlation Plots program (line 14; imported, as specified in line 6, from \texttt{tools/xpcs\_plot.py}\footnote{\url{https://github.com/globus-gladier/gladier-patterns-examples-2022/blob/main/tools/xpcs_plot.py}}).
\end{enumerate}

Subsequent statements in \texttt{simple\_xpcs\_client.py} configure various parameters, including the UUIDs that identify the funcX endpoint that is to be used to run the {\sc{Compute}} tasks (\texttt{analysis\_computer\_funcx\_id}) and the source and destination Globus collections (\texttt{instrument\_computer\_collection\_id} and \texttt{analysis\_computer\_collection\_id}) for the {\sc{Transfer}} task.
A value is already provided for \texttt{instrument\_computer\_collection\_id}, the source of the data to be processed. Normally, this would be a storage system at the XPCS instrument, but it is configured in \texttt{simple\_xpcs\_client.py} to be a collection that we have established to store XPCS test data. Values are not provided, on the other hand, for \texttt{analysis\_computer\_collection\_id} or \texttt{analysis\_computer\_funcx\_id}. We will show in the next steps how to configure these on your personal computer. 

When first initialized, this code generates a flow definition and registers it with the Globus Flows service. 
It also registers the two funcX tools, \texttt{BoostCorr} and \texttt{MakeCorrPlots}, with the funcX service. 
Subsequent invocations reuse the registered flow and functions.


The \texttt{simple\_xpcs\_client.py} application is easy to run on your own computer. The steps are:

\begin{enumerate}
    \item 
    \textbf{Establish the destination Globus collection}.
    As noted, the application needs a value for
      \texttt{analysis\_computer\_collection\_id}, the identifier of a Globus collection accessible from the computer on which analysis tasks are to be executed. If no such collection is accessible to us, we can create a new collection by installing and configuring Globus Connect Personal software, as described online for Linux, MacOS, and Windows computers.\footnote{\url{https://www.globus.org/globus-connect-personal}} We then record the UUID for the collection by setting it as the value of \texttt{analysis\_computer\_collection\_id} in the \texttt{simple\_xpcs\_client.py} application.

    \item 
    \textbf{Specify the funcX endpoint}.
    The application also needs a value for \texttt{analysis\_computer\_funcx\_id}, the identifier of the funcX endpoint where analysis tasks are to be executed. If no such endpoint is accessible to us, we can create a new funcX endpoint on our personal computer by installing and configuring the funcX software, as described in the repository's \texttt{README.md} file.\footnoteref{repo:xpcs} We then record the UUID for the funcX endpoint by setting it as the value of \texttt{analysis\_computer\_funcx\_id} in the \texttt{simple\_xpcs\_client.py} application.

    \item 
    \textbf{Configure execution environment on compute endpoint(s)}.
    The funcX system that we use to implement {\sc{Compute}} actions can run any Python functions or containerized programs invokable from Python that have been registered with the funcX service. We install programs that cannot be thus registered (e.g., a non-containerized application) manually prior to use, so that they may be invoked by {\sc{Compute}} actions during flow execution. Here we installed four such programs: the 
    XPCS Boost correlation analysis tool,\footnote{\label{fn:boost_corr}\url{https://github.com/AZjk/boost_corr}}
    CUDA Toolkit,\footnote{\label{fn:cuda}\url{https://developer.nvidia.com/cuda-toolkit}} PyTorch,\footnote{\label{fn:torch}\url{https://pypi.org/project/torch/}} and
    Gladier XPCS repository,\footnote{\label{fn:xpcs}\url{https://github.com/globus-gladier/gladier-xpcs}} which includes custom plotting modules. 
    
    \item
    \textbf{Run the application}.
    We start the flow by executing the supplied \texttt{simple\_xpcs\_client.py}. When first invoked, the user is prompted to login and consent to the flow accessing the Transfer and funcX services. The application provides a link to the Globus Flows service where the flow can be monitored.
    
\end{enumerate}

\subsection{Other simplified applications}

\noindent
The \textbf{simplified SSX application} (specifically, a simplified version of the SSX-Stills flow described in the paper\footnote{\url{https://github.com/globus-gladier/gladier-patterns-examples-2022/blob/main/simple_ssx_client.py}}) implements a flow with four steps: a \textsc{Transfer} from instrument to analysis computer followed by three \textsc{Compute} steps that create a  Phil-format\footnote{\url{http://cctbx.sourceforge.net/libtbx_phil.html}} input file for the DIALS Stills application, run DIALS Stills, and run DIALS unit\_cell\_histogram, respectively.

\vspace{1ex}
\noindent
The \textbf{simplified HEDM application}\footnote{\url{https://github.com/globus-gladier/gladier-patterns-examples-2022/blob/main/simple_hedm_client.py}}
implements a flow with two steps: a \textsc{Transfer} from instrument to computer and a \textsc{Compute} step that runs a supplied shell script.

\vspace{1ex}
\noindent
The \textbf{simplified BraggNN application}\footnote{\url{https://github.com/globus-gladier/gladier-patterns-examples-2022/blob/main/simple_braggnn_client.py}} implements a flow with two steps: a \textsc{Transfer} from instrument to computer, and a \textsc{Compute} step that runs a supplied shell script.

\vspace{1ex}
\noindent
The \textbf{simplified Ptychography application}\footnote{\url{https://github.com/globus-gladier/gladier-patterns-examples-2022/blob/main/simple_ptycho_client.py}}
implements a flow with three steps: a \textsc{Transfer} from instrument to computer, and two \textsc{Compute} steps that run a supplied shell script and the \texttt{ptychodus\_plot} tool,\footnote{\url{https://github.com/globus-gladier/gladier-patterns-examples-2022/blob/main/tools/ptychodus_plot.py}} respectively.

\section{The full applications and flows described in the paper}\label{sec:SIflows}

\noindent
The source code for the five applications described in this paper is on GitHub. We provide pointers to each application's source code and notes on the steps involved in running each.
These production applications differ from the simplified applications described in \ref{sec:SI-xpcs} in various ways. In particular, they:
\begin{itemize}
    \item 
    define separate funcX endpoints for non-compute-intensive and compute-intensive \textsc{Compute} tasks, respectively (on an HPC system, these will typically correspond to a login node vs.\ compute nodes); and
    \item 
    publish descriptive metadata plus data references to a Globus Search catalog, and establish an associated interactive data portal, so that users can browse, search, and access flow products.
\end{itemize}

\subsection{The XPCS application}

\noindent
Code and documentation on GitHub\footnoteref{fn:xpcs} support the processing of XPCS data generated at the 8-ID beamline of the Advanced Photon Source (APS).
The generation of spectroscopy data at 8-ID triggers a flow that transfers data from 8-ID to ALCF for analysis, metadata extraction, and visualization, and then publishes the processed data to an ALCF Community Data Co-Op\footnote{\url{https://acdc.alcf.anl.gov}} portal.  

The Python program \texttt{flow\_boost.py}\footnote{\url{https://github.com/globus-gladier/gladier_xpcs/flows/flow_boost.py}} implements the flow described in the paper, with the addition of a step~5 to preallocate nodes on the HPC resource, an optimization that can accelerate flow start. Some notes about how to configure the flow to run:

\begin{enumerate}
    \item 
    \textbf{Configure infrastructure}:
    The XPCS flow involves {\sc{Transfer}}, {\sc{Compute}}, and {\sc{Search}} actions.

    \begin{itemize}
        \item 
            As the flow involves {\sc{Transfer}} actions, we must ensure that \textbf{Globus collections} are in place wherever data are to be accessed: in this case, the APS 8-ID beamline and ALCF Eagle storage systems.
            As Globus collections are already deployed in both locations as part of their regular infrastructure, no action was required.
    
        \item
            As the flow involves {\sc{Compute}} actions, we must ensure that \textbf{funcX endpoints} are deployed wherever computation is to be performed: in this case, the ALCF Theta computer. The endpoint must also be configured to interface with the batch scheduler to appropriately acquire nodes. Here, we define a Cobalt configuration using the example in the funcX documentation.\footnote{\url{https://funcx.readthedocs.io/en/latest/endpoints.html\#theta-alcf}}
    
        \item 
            As the flow involves {\sc{Search}} actions, we must ensure that a \textbf{Globus Search index} has been provisioned and a data portal deployed and customized to visualise search records. The XPCS search index was created via the Globus CLI.\footnote{ \url{https://docs.globus.org/cli/reference/search_index_create/}} 
            The XPCS data portal was implemented by using the Django Globus Portal Framework,\footnote{\url{https://github.com/globus/django-globus-portal-framework}} with customization to display specific metadata, facets, and images. The portal implementation and installation instructions are on Github.\footnote{\url{https://github.com/globus-gladier/gladier-xpcs/tree/main/xpcs_portal}}
        
    \end{itemize}

    \item
        \textbf{Configure execution environment on compute endpoint(s)}.
        As with the simplified XPCS application, we install four programs that cannot be registered automatically with the funcX service: the XPCS Boost correlation analysis tool,\footnoteref{fn:boost_corr} CUDA Toolkit,\footnoteref{fn:cuda} PyTorch,\footnoteref{fn:torch} and the Gladier XPCS repository,\footnoteref{fn:xpcs} which includes custom plotting modules. 

    \item
    \textbf{Configure flow triggers}.
    A trigger may be configured to invoke an instance of a flow in response to data being generated.
    In this example, instances of the flow are initiated by the APS Data Management System,\footnote{S.~Veseli, N.~Schwarz, C.~Schmitz. ``APS data management system,'' Journal of Synchrotron Radiation 25(5):1574-1580, 2018, \url{https://doi.org/10.1107/S1600577518010056}.}
    which copies each batch of new images, as they are acquired, from the instrument to storage accessible by Globus Transfer, and then starts an instance of the flow.

\end{enumerate}

\subsection{The SSX application}

\noindent
The code for the full SSX application is on GitHub.\footnote{\url{https://github.com/globus-gladier/gladier-kanzus}}

\subsection{The HEDM application}

\noindent
The code for the full HEDM application is on GitHub.\footnote{\url{https://github.com/globus-gladier/gladier-hedm}}

\subsection{The BraggNN application}

\noindent
The code for the full BraggNN application is on GitHub.\footnote{\url{https://github.com/lzhengchun/nnTrainFlow}}

\subsection{The Ptychography application}

\noindent
The code for the full Ptychography application is on GitHub.\footnote{\url{https://github.com/globus-gladier/gladier-ptycho}}

\section{Other software referenced in, or relevant to, the paper}\label{sec:SIsoftware}

\noindent
The \textbf{Gladier Toolkit}\footnote{\url{https://github.com/globus-gladier/gladier}}\footnote{\url{https://gladier.readthedocs.io}} (see body of paper) is designed to accelerate and simplify the implementation of new scientific flows for experimental facilities. It provides a Pythonic interface for defining Globus flows and for managing the registration and caching of flows and of funcX functions. 

\vspace{1ex}
\noindent
The supporting \textbf{Gladier Tools}\footnote{\url{https://github.com/globus-gladier/gladier-tools}}\footnote{\url{https://gladier.readthedocs.io/en/latest/gladier_tools}} package utility tools that can be incorporated into a flow, such as Transfer and Publication. Tools in this repository are intended to be general purpose and reusable.

\vspace{1ex}
\noindent
The \textbf{Globus Python SDK}\footnote{\url{https://github.com/globus/globus-sdk-python}}\footnote{\url{https://globus-sdk-python.readthedocs.io}} provides a convenient Pythonic interface to Globus web APIs, including the Globus Transfer API and Globus Auth API. 
It is used extensively by the Gladier Toolkit and tools.

\vspace{1ex}
\noindent
The \textbf{Globus Automate CLI and SDK}\footnote{\url{https://github.com/globus/globus-automate-client}}\footnote{\url{https://globus-automate-client.readthedocs.io}} provides a command line interface (CLI) and Python software development kit (SDK) for working with Globus automation services, primarily Globus Flows, any service implementing the Globus Action Provider interface, and Globus Queues.

\vspace{1ex}
\noindent
The \textbf{Globus Sample Data Portal}\footnote{\url{https://github.com/globus/globus-sample-data-portal}}\footnote{\url{https://docs.globus.org/modern-research-data-portal}} implements a simple Web app framework that illustrates how to build a data portal, such as those created for the example applications presented in this paper, by using Globus services.
 
\vspace{1ex}
\noindent       
The \textbf{Django Globus Portal Framework}\footnote{\url{https://github.com/globus/django-globus-portal-framework}}\footnote{\url{https://django-globus-portal-framework.readthedocs.io}} provides a modular framework for building Globus-based data portals. It provides utilities for quickly building a data portal around a Globus Search index, using Globus Auth to secure access to data.


\end{document}